\documentclass[%
reprint,
nofootinbib,
amsmath,amssymb,
aps,
prd,
floatfix
]{revtex4-1}

\usepackage{caption}
\usepackage{subcaption}
\usepackage{multirow}
\usepackage{graphicx}  
\usepackage{amsmath}   
\usepackage{amssymb}   
\usepackage{bm} 
\usepackage{dcolumn}
\usepackage{color}
\usepackage{mathrsfs}
\usepackage{amsfonts}
\usepackage{varioref}
\usepackage{mathrsfs}
\usepackage{graphicx}
\usepackage{latexsym}
\usepackage{amsmath}
\usepackage{amssymb}
\usepackage{textcomp}
\usepackage{amsbsy}
\usepackage{graphics}
\usepackage{epstopdf}
\usepackage{color}
\usepackage{enumitem}
\usepackage{array,multirow} 
\usepackage{float}
\usepackage{extarrows}
\usepackage{url}
\usepackage{caption}
\usepackage{orcidlink}

\usepackage{booktabs, adjustbox}

\allowdisplaybreaks

\input epsf

\usepackage{hyperref}
\hypersetup{
    colorlinks=true,
    citecolor=blue,
    linkcolor=blue,
    filecolor=magenta,      
    urlcolor=magenta,
}

\newcommand{\Omegab}{\Omega_\text{b}}
\newcommand{\Omegac}{\Omega_\text{c}}
\newcommand{\Omegamzero}{\Omega_\text{m0}}
\newcommand{\Omegarzero}{\Omega_\text{r0}}

\newcommand{\OmegaDE}{\Omega_\text{DE}}

\newcommand{\omegab}{\omega_\text{b}}
\newcommand{\omegabc}{\omega_\text{bc}}

\newcommand{\Neff}{N_\text{eff}}

\newcommand{\rhoDE}{\rho_\text{DE}}
\newcommand{\Mpl}{M_\text{Pl}}
\begin{document}

\tolerance=5000

\title{Holographic dark energy from generalized entropies and extended infrared cut-offs}

\author{Ke Chen$^{1}$\,\orcidlink{0009-0001-2207-0128} \thanks{kchen@ice.csic.es},
Simone~D'Onofrio$^{1}$\,\orcidlink{0000-0001-7925-3905} \thanks{donofrio@ice.csic.es},
Sergei~D.~Odintsov$^{1,3}$,\orcidlink{0000-0002-3529-7030}
\thanks{odintsov@ice.csic.es},
Tanmoy~Paul$^{2}$\,\orcidlink{0000-0002-0854-3427}
\thanks{tanmoy.paul@visva-bharati.ac.in}}
\affiliation{
$^{1)}$ Institute of Space Sciences (ICE, CSIC) C. Can Magrans s/n, 08193 Barcelona, Spain\\
$^{2)}$ Department of Physics, Visva-Bharati University, Santiniketan 731235, West Bengal, India\\
$^{3)}$ ICREA, Passeig Luis Companys, 23, 08010 Barcelona, Spain}

\tolerance=5000

\begin{abstract}
We study holographic dark energy models constructed from generalized entropy
functionals and local infrared cut-offs. Three entropies (Barrow and the three-
and five-parameter generalized entropies) are each combined with the
Granda--Oliveros (GO) cut-off and two extensions, one adding a $\gamma H^{-2}$ term
and one a $\gamma\,\ddot{H}/H$ term, the latter raising the dark-energy evolution
equation to second order. The nine resulting models are constrained against DESI DR2
BAO, Pantheon+ and a Planck compressed CMB likelihood through an MCMC analysis,
and compared using the Akaike and Bayesian information criteria. The Hubble constant
is stable across all models. The $H^{-2}$ term is unconstrained in every case, as it
becomes relevant only in the far future, and never improves on the GO cut-off. The
$\ddot{H}/H$ term, instead, is tightly constrained for Barrow and $S_5$, where it
improves the fit substantially, but not for $S_3$. The best fit is obtained for Barrow
entropy with the $\ddot{H}/H$ cut-off, yet the Barrow exponent remains compatible with
zero in all cases. The data thus favour a higher-order infrared cut-off, while showing
no significant evidence for a departure from the Bekenstein--Hawking area law.
We also investigate the dark energy equation of state (EoS) of $S_3$ and $S_5$, discovering quintessence-like behavior and phantom crossing.
\end{abstract}

\maketitle

\section{Introduction}
 
The physical origin of the late-time accelerated expansion of the Universe, first
established by Type Ia supernova observations~\cite{Riess:1998cb,Perlmutter:1999np},
remains one of the central open problems of modern cosmology. The concordance
$\Lambda$CDM model, in which acceleration is driven by a cosmological constant,
provides an excellent fit to current data~\cite{Planck_2020}, yet it is challenged
by the fine-tuning and coincidence problems~\cite{Weinberg:1988cp} and by the persistent
discrepancy between early- and late-time determinations of
$H_0$~\cite{Riess:2021jrx,DiValentino:2021izs}. Together with the recent DESI hints of
an evolving dark energy component~\cite{DESI:2024mwx,DESI:2025zgx}, this motivates the
study of dynamical dark energy alternatives and modified gravity
(see, e.g.,~\cite{Bamba:2012cp,Peebles:2002gy,Nojiri:2006ri,Capozziello:2011et,Nojiri:2017ncd}
for reviews).
 
Holographic dark energy (HDE) offers a particularly appealing route, since it ties
the dark energy density to fundamental considerations rooted in the holographic
principle~\cite{Cohen:1998zx}. The construction rests on the idea that the vacuum
energy inside a region of size $L$ cannot exceed that of a black hole of the same
size, which leads to an energy density $\rho_{\rm DE}\propto M_{\rm Pl}^2 L^{-2}$.
The resulting cosmology depends crucially on the choice of the infrared (IR) cut-off
$L$~\cite{Li:2004rb,Hsu:2004ri}, and an extensive body of work has explored
prescriptions ranging from the Hubble and particle horizons to local, causal
constructions built from $H$ and its derivatives~\cite{Huang:2004ai,Pavon:2005yx,
Elizalde:2005ju,Granda:2008dk,Granda:2009fh,Nojiri:2005pu,Nojiri:2017opc,Wang:2016och,
Nojiri:2020wmh}; in fact, every such model can be viewed as a particular case of the
Nojiri--Odintsov (NO) generalized HDE~\cite{Nojiri:2005pu,Nojiri:2017opc,Nojiri:2021iko}.
Among the local prescriptions, the Granda--Oliveros (GO)
cut-off, $L^{-2}=\alpha H^2+\beta\dot H$, is especially attractive: it depends only
on local, present-epoch quantities, avoids the causality issues of the future event
horizon, and alleviates the coincidence problem.
 
A complementary generalisation replaces the standard Bekenstein--Hawking area
law~\cite{Bekenstein:1973ur,Hawking:1975vcx} by a generalized entropy
functional~\cite{Nojiri:2022nof,Nojiri:2021jxf,Nojiri:2022sfd,Tsallis:1987eu,Nojiri:2019skr}
(for general reviews of entropic cosmology, see~\cite{Nojiri:2024eac,Saridakis:2026rev}),
leading, e.g., to Tsallis~\cite{Tavayef:2018xwx,DAgostino:2019wko},
R\'enyi~\cite{Moradpour:2018ivi} and Kaniadakis~\cite{Drepanou:2021jiv} holographic dark
energy. Barrow entropy~\cite{Barrow:2020tzx}, motivated by a
fractal deformation of the horizon, is also known as the Sharma--Mittal entropy,
which was introduced earlier in a different context~\cite{Sharma:1975}. It has been shown to yield a richer dark energy
phenomenology~\cite{Saridakis:2020zol,Anagnostopoulos:2020ctz,Leon:2021wyx}, and
multi-parameter families such as the three-parameter ($S_3$) and five-parameter
($S_5$) generalized entropies~\cite{Nojiri:2022nof,Nojiri:2022aof,Nojiri:2022dkr,Odintsov:2022qnn}
provide unifying frameworks that interpolate between many known entropic scenarios.
The combination of generalized entropies with the GO cut-off has attracted growing
attention in the light of recent data~\cite{Oliveros:2022biu,Tyagi:2024cqp,
Mahmoudifard:2024epjc,Yarahmadi:2024mnras,Luciano:2025jcap,Luciano:2026jheap,
Chattopadhyay:2025epjc}. What is still missing, however, is a
systematic and homogeneous confrontation of \emph{several} entropy functionals with
\emph{several} IR cut-off prescriptions against a single, state-of-the-art dataset,
analyzed with a common Bayesian pipeline.
 
In this work we fill this gap by studying a $3\times 3$ grid of nine models: three
generalized entropy functionals---Barrow, $S_3$ and $S_5$---each combined with three
local IR cut-offs, with the standard Bekenstein--Hawking entropy recovered as the
$\Delta\to 0$ baseline. The three cut-offs are the GO scale and two extensions
introduced here: cut-off 2, which adds a term $\gamma H^{-2}$ that switches on only in
the far-future, low-$H$ regime, and cut-off 3, which adds a term $\gamma\,\ddot H/H$
and thereby promotes the evolution equation for $\Omega_{\rm DE}(z)$ to second order.
To our knowledge the $H^{-2}$ combination has not been considered before, while the
$\ddot H/H$ extension has so far been examined only with cosmic-chronometer
data~\cite{Singh:2026ihe}; the joint-dataset analysis presented here is therefore
largely unexplored territory.
 
We constrain all nine models against a joint dataset combining DESI DR2 baryon
acoustic oscillation measurements~\cite{DESI:2025zgx}, the Pantheon+ Type Ia
supernova compilation~\cite{Brout_2022}, and a Planck 2018 compressed CMB
likelihood~\cite{Planck_2020,Lemos:2023xhs}. Posterior distributions are obtained
with the \texttt{emcee} sampler~\cite{Foreman-Mackey2019}, and the models are ranked
using the Akaike and Bayesian information criteria. Our aims are threefold: to
determine which entropy--cut-off combinations are observationally viable; to assess
whether the extended cut-offs provide genuine dynamical leverage beyond GO or merely
add unconstrained parameters; and to extract the physical lessons encoded in the
parameter degeneracies and information-criteria penalties.
 
The paper is organized as follows. Section~\ref{sec:BH} introduces the HDE
construction based on Bekenstein--Hawking entropy and derives the evolution
equations for the GO cut-off and cut-offs 2 and 3. Sections~\ref{sec:Barrow},
\ref{sec:S3} and~\ref{sec:S5} extend the framework to the Barrow, $S_3$ and $S_5$
entropies, highlighting the qualitative change of structure---most notably the
appearance of a second-order (cut-off 3) or higher-order system---that arises for each.
Section~\ref{sec:obs_data} describes the observational datasets and the likelihood
construction, and details the MCMC methodology and priors.
Section~\ref{sec:results} presents the constraints and model comparison for all nine
combinations, organized by entropy sector, together with the cross-cutting patterns.
Our conclusions are summarized in Section~\ref{sec:conclusions}.

\section{Holographic Dark Energy Based on Bekenstein-Hawking Entropy}\label{sec:BH}
Holographic dark energy (HDE) models provide an interesting framework to derive
dynamical DE in agreement with the holographic principle. Firstly introduced in
\cite{Cohen:1998zx}, these models are based on the principle that the energy inside
a region of size $L$ cannot exceed that of a black hole of the same size,
\begin{equation}
    L^3 \rhoDE \leq L \Mpl^2 \ .
\end{equation}
This equation relates the long-distance cut-off (l.h.s.) to the short-scale one
(r.h.s.). It will be useful to rewrite the inequality as
\begin{equation}
    \rhoDE = 3 C^2 \Mpl^2 L^{-2} \ ,
\end{equation}
where $C$ is a positive constant.

The choice of the infrared cut-off $L$ is crucial. Among the possibilities we have
the size of the current universe (the Hubble scale $L=1/H$), the particle horizon
$L_p$ and the future event horizon $L_f$. In \cite{Nojiri:2005pu}, it has been shown
that it can take the general form
\begin{equation}
    L = L \left(L_p,\Dot{L}_p,\Ddot{L}_p,\dots,L_f,\Dot{L}_f,\Ddot{L}_f,\dots,
    a, H, \Dot{H}, \Ddot{H}, \dots\right) \ ,
\end{equation}
a framework subsequently formalized in covariant form in \cite{Nojiri:2017opc}.
It was shown that any HDE represents a particular case of this Nojiri--Odintsov (NO)
generalized HDE~\cite{Nojiri:2021iko,Nojiri:2022cez}. In the present work we consider the following three choices of infrared
cut-off, each extending the previous one by an additional term. For each model, the
cosmological evolution is fully determined by combining the definition of the
fractional dark energy density
\begin{equation}
    \OmegaDE(z) = \frac{\rhoDE(z)}{3\Mpl^2 H^2(z)}
\end{equation}
with the Friedmann equation, which we write as
\begin{equation}\label{eq:Hubble}
    H(z) = H_0 \sqrt{\frac{\Omegamzero(1+z)^3 + \Omegarzero(1+z)^4}{1 - \OmegaDE(z)}} \ .
\end{equation}
Using $d/dt = -H(1+z)\,d/dz$, the definition of $\OmegaDE$ for each cut-off can
be recast as an ODE, which is then solved numerically subject to the boundary
condition
\begin{equation}\label{eq:BC}
    \OmegaDE(0) = 1 - \Omegamzero - \Omegarzero \ ,
\end{equation}
imposed at $z=0$. It is convenient to introduce
\begin{align}
    \Sigma(z)&\equiv\Omegamzero+\Omegarzero(1+z)\,,\nonumber\\
    \mathcal{E}(z)&\equiv\frac{d\ln H^2}{dz}=\frac{3}{1+z}+\frac{\Omegarzero}{\Sigma}+\frac{\OmegaDE'}{1-\OmegaDE}\,,\label{eq:Sigma_E}
\end{align}
where the last equality follows from Eq.~\eqref{eq:Hubble}. In terms of $\mathcal{E}$,
the time derivatives of the Hubble rate read
\begin{equation}\label{eq:Hdots}
    \frac{\dot H}{H^2}=-\frac{1+z}{2}\,\mathcal{E}\,,\qquad
    \frac{\ddot H}{H^3}=\frac{1+z}{2}\Bigl[\mathcal{E}+(1+z)\bigl(\mathcal{E}^2+\mathcal{E}'\bigr)\Bigr]\,.
\end{equation}

\subsection{GO cut-off}

The Granda--Oliveros (GO) cut-off reads
\begin{equation}
    L_{\rm GO} = \left(\alpha H^2 + \beta \Dot{H}\right)^{-1/2} \ ,
    \label{eq:GO}
\end{equation}
where $\alpha$ and $\beta$ are dimensionless constants. Originally proposed
in \cite{Granda:2008dk,Granda:2009fh}, this cut-off generalises the holographic
Ricci scale~\cite{Cai:2008wt,Gao:2007ep} by allowing $\alpha$ and $\beta$ to be free
parameters. Its key virtues are that it depends only on local, present-epoch
quantities, thereby avoiding the causality problem associated with the future event
horizon, and that it alleviates the coincidence problem. The GO cut-off arises as a
special case of the Nojiri--Odintsov general framework~\cite{Nojiri:2005pu,Nojiri:2017opc,Nojiri:2021iko,Nojiri:2022cez}, and has been extensively studied in the context of Barrow
holographic dark energy
\cite{Oliveros:2022biu,Mahmoudifard:2024epjc,Luciano:2025jcap,Luciano:2026jheap}.

With the GO cut-off, the standard HDE density $\rhoDE = 3C^2\Mpl^2 L_{\rm GO}^{-2}$
gives $\OmegaDE = C^2\left(\alpha + \beta\Dot{H}/H^2\right)$. Using Eqs.~\eqref{eq:Sigma_E}
and~\eqref{eq:Hdots}, the fractional density satisfies the first-order ODE
\begin{align}\label{eq:ODE_GO}
    \OmegaDE'(z) ={}& \left(1 - \OmegaDE\right)\biggl[\frac{2\left(\alpha - \OmegaDE/C^2\right)}{\beta\,(1+z)}\nonumber\\
    &- \frac{3}{1+z} - \frac{\Omegarzero}{\Sigma}\biggr] ,
\end{align}
which, together with Eq.~\eqref{eq:Hubble} and the boundary condition
\eqref{eq:BC}, constitutes a closed system for the cosmological evolution.

\subsection{Cut-off 2}

The second cut-off we consider (hereafter cut-off 2) is
\begin{equation}
    L_{(2)} = \left(\alpha H^2 + \beta \Dot{H}
    + \gamma H^{-2}\right)^{-1/2} \ ,
    \label{eq:EGOI}
\end{equation}
where $\gamma$ weights a term inversely proportional to $H^2$; for all three terms
to have dimension $[\text{time}]^{-2}$, $\gamma$ carries dimension $[\text{time}]^{-4}$.
The $\gamma H^{-2}$ contribution is suppressed at early
times when $H$ is large and becomes relevant in the late-time low-$H$ regime,
providing a complementary extension to the GO cut-off compared to the positive
higher-power generalisation $\alpha H^2 + \beta\Dot{H} + \gamma H^4$ studied in
\cite{Chattopadhyay:2025epjc} within the Nojiri--Odintsov framework. To our
knowledge, the specific combination with an $H^{-2}$ term has not been previously
considered in the literature and represents a novel proposal of the present work.

The additional term introduces an explicit dependence on $H(z)$ in the ODE, which
now reads
\begin{align}\label{eq:ODE_EGOI}
    \OmegaDE'(z) ={}& \left(1 - \OmegaDE\right)\biggl[\frac{2\left(\alpha + \gamma H^{-4} - \OmegaDE/C^2\right)}{\beta\,(1+z)}\nonumber\\
    &- \frac{3}{1+z} - \frac{\Omegarzero}{\Sigma}\biggr] ,
\end{align}
where $H(z)$ is given by Eq.~\eqref{eq:Hubble}. Together with the boundary
condition \eqref{eq:BC}, this provides a closed system analogous to the GO case.

\subsection{Cut-off 3}

The third cut-off (hereafter cut-off 3) is
\begin{equation}
    L_{(3)} = \left(\alpha H^2 + \beta \Dot{H}
    + \gamma \frac{\Ddot{H}}{H}\right)^{-1/2} \ ,
    \label{eq:EGOII}
\end{equation}
where $\gamma$ now weights the contribution of the second time derivative of the
Hubble parameter. The ratio $\Ddot{H}/H$ has dimension $[\text{time}]^{-2}$, so
that $\gamma$ is dimensionless, like $\alpha$ and $\beta$. This higher-order extension belongs to the
general class of derivative-based cut-offs encompassed by the Nojiri--Odintsov
framework \cite{Nojiri:2005pu,Nojiri:2017opc}. Such terms also arise naturally when
modified-gravity cosmologies are recast in holographic form: in $F(R)$ gravity, for
instance, the effective dark energy density depends on $\dot R$, and hence on
$\ddot H$~\cite{DOnofrio:2023jgc}. A related model in which the IR
cut-off depends on $H$, $\Dot{H}$, and $\Ddot{H}$ has recently been studied in
\cite{Singh:2026ihe} in the context of a long-freeze cosmology, where the
$\Ddot{H}$ contribution is interpreted as an effective entropic inertia of the
holographic vacuum. Our cut-off 3 differs in that it embeds the
$\Ddot{H}$ term within the local GO structure, made dimensionally homogeneous through the factor $H^{-1}$.

The presence of $\Ddot{H}$, which translates into $\OmegaDE''(z)$ upon changing
variable to redshift, makes this model qualitatively different from the
previous two. Inserting Eq.~\eqref{eq:Hdots} into
$\OmegaDE = C^2\bigl(\alpha + \beta\Dot{H}/H^2 + \gamma\Ddot{H}/H^3\bigr)$ gives
\begin{equation}\label{eq:Eprime}
    \mathcal{E}' = \frac{2\left(\OmegaDE/C^2-\alpha\right)}{\gamma\,(1+z)^2}
    + \frac{\beta-\gamma}{\gamma\,(1+z)}\,\mathcal{E} - \mathcal{E}^2 \ ,
\end{equation}
and differentiating Eq.~\eqref{eq:Sigma_E} turns this into the \emph{second-order} ODE
\begin{equation}\label{eq:ODE_EGOII}
    \OmegaDE''(z) = \left(1-\OmegaDE\right)\left[\mathcal{E}' + \frac{3}{(1+z)^2} + \frac{\Omegarzero^2}{\Sigma^2}\right] - \frac{\OmegaDE'^{\,2}}{1-\OmegaDE} \ .
\end{equation}
The boundary condition \eqref{eq:BC} fixes $\OmegaDE(0)$, but the system now
requires a second condition on $\OmegaDE'(0)$. Unlike the GO and cut-off 2 cases, here $\OmegaDE'(0)$ is not fixed by the model equations alone and is
therefore treated as a free parameter, sampled alongside the other model
parameters in the MCMC analysis.

\section{Holographic Dark Energy Based on Barrow Entropy}\label{sec:Barrow}

The idea of combining the holographic dark energy (HDE) formalism with generalized entropy functionals~\cite{Nojiri:2022nof,Nojiri:2021jxf,Nojiri:2022sfd} has attracted considerable attention as a natural extension of the standard HDE paradigm \cite{Cohen:1998zx, Li:2004rb, Wang:2016och, Nojiri:2021iko, Saridakis:2026rev}. In the standard approach, the dark energy density is determined by an infrared (IR) cut-off $L$ through the relation $\rho_\mathrm{DE} \propto M_\mathrm{Pl}^2 L^{-2}$, which follows from applying the holographic bound to the vacuum energy \cite{Hsu:2004ri, Li:2004rb}. Generalizing the underlying entropy functional modifies this relation and leads to richer cosmological dynamics.

The Barrow entropy \cite{Barrow:2020tzx} arises from the proposal that quantum-gravitational effects endow the black hole horizon with a fractal, intricate structure, leading to a modified area--entropy relation
\begin{equation}\label{S_Barrow}
    S_B = \left(\frac{A}{A_0}\right)^{1+\frac{\Delta}{2}} = \left(\frac{L}{L_0}\right)^{2+\Delta} \ ,
\end{equation}
where $\Delta \in [0,1]$ is the Barrow deformation parameter quantifying the degree of fractal complexity: $\Delta = 0$ recovers the standard Bekenstein--Hawking entropy, while $\Delta = 1$ corresponds to the maximally complex case. This entropy has been extensively studied in the context of modified cosmology and thermodynamics \cite{Saridakis:2020zol, Anagnostopoulos:2020ctz, Leon:2021wyx}.

Within the HDE framework, the IR cut-off $L$ is linked to an effective dark energy density via
\begin{equation}\label{rhode_Barrow}
    \rho_\mathrm{DE} = S(L)\, L^{-4} = B\, L^{\Delta - 2} \ ,
\end{equation}
where the last equality follows from substituting the Barrow entropy \eqref{S_Barrow}, and the normalization constant
\begin{equation}
    B = 3 C^2 M_\mathrm{Pl}^2
\end{equation}
encodes the HDE free parameter $C > 0$. This mixing of generalized entropy and the HDE formalism was introduced in \cite{Oliveros:2022biu} for the specific combination of Barrow entropy and the Granda-Oliveros (GO) cut-off, and further studied in \cite{Tyagi:2024cqp} where the two formalisms are applied separately to late-time cosmology.

\subsection{GO cut-off}

For the IR cut-off we adopt the Granda--Oliveros proposal \cite{Granda:2008dk, Granda:2009fh}
\begin{equation}
    L = \left(\alpha H^2 + \beta \dot{H}\right)^{-1/2} \ ,
\end{equation}
which is a local, covariant definition depending on the Hubble rate and its time derivative. This choice avoids the causality issues associated with the event horizon cut-off \cite{Hsu:2004ri} and has been shown to yield a consistent cosmological evolution with good observational fits \cite{Granda:2008dk, Granda:2009fh, Oliveros:2022biu}. With this cut-off the dark energy density becomes
\begin{equation}\label{rho_Barrow_GO}
    \rho_\mathrm{DE} = 3 C^2 M_\mathrm{Pl}^2 \left(\alpha H^2 + \beta \dot{H}\right)^{1+\Delta/2} \ .
\end{equation}
Note that for $\Delta = 0$ this reduces to the standard GO holographic dark energy density $\rho_\mathrm{DE} = 3C^2 M_\mathrm{Pl}^2 (\alpha H^2 + \beta \dot{H})$, consistently with earlier results \cite{Granda:2008dk}.

The background dynamics is governed by the first Friedmann equation
\begin{equation}
    3 M_\mathrm{Pl}^2 H^2 = \rho_m + \rho_r + \rho_\mathrm{DE} \ .
\end{equation}
Defining the fractional energy densities
\begin{equation}\label{fed}
    \Omega_i(z) \equiv \frac{\rho_i(z)}{3 M_\mathrm{Pl}^2 H^2(z)} \ ,
\end{equation}
the Hubble parameter is given by Eq.~\eqref{eq:Hubble}, which we rewrite as
\begin{equation}\label{H_z}
    H(z) = H_0 \sqrt{\frac{(1+z)^3\,\Sigma(z)}{1 - \Omega_\mathrm{DE}(z)}} \ ,
\end{equation}
while the fractional dark energy density takes the form
\begin{equation}
    \Omega_\mathrm{DE}(z) = \frac{C^2}{H^2} \left(\alpha H^2 + \beta \dot{H}\right)^{1+\Delta/2} \ .
\end{equation}
Inverting this relation, the holographic constraint fixes the value of $L^{-2}$ to
\begin{equation}\label{def_F}
    \mathcal{F} \;\equiv\; \left(\frac{\Omega_\mathrm{DE}\, H^2}{C^2}\right)^{\!\frac{2}{2+\Delta}} \ ,
\end{equation}
so that, using Eqs.~\eqref{eq:Sigma_E} and~\eqref{eq:Hdots}, one obtains a single first-order ODE for $\Omega_\mathrm{DE}(z)$:
\begin{align}\label{ODE_Barrow_GO}
    \Omega_\mathrm{DE}'(z) ={}& \left(1 - \Omega_\mathrm{DE}\right)\biggl[\frac{2\left(\alpha - \mathcal{F}/H^2\right)}{\beta\,(1+z)}\nonumber\\
    &- \frac{3}{1+z} - \frac{\Omega_{r0}}{\Sigma}\biggr] .
\end{align}
For $\Delta = 0$ one has $\mathcal{F}/H^2 = \Omega_\mathrm{DE}/C^2$ and Eq.~\eqref{eq:ODE_GO} is recovered. Together with Eq.~\eqref{H_z}, this constitutes a closed system that fully determines the cosmological evolution of the model. The initial condition $\Omega_\mathrm{DE}(0) = 1 - \Omega_{m0} - \Omega_{r0}$ is imposed at $z=0$, and the system is integrated numerically to reconstruct the expansion history.

\subsection{Cut-off 2}

Cut-off 2 supplements the GO definition with a term inversely proportional to $H^2$:
\begin{equation}\label{EGO-I}
    L^{-2} = \alpha H^2 + \beta\,\dot{H} + \gamma\, H^{-2}\,.
\end{equation}
Since no time derivatives beyond $\dot{H}$ appear, the derivation proceeds identically to the GO case, and the evolution equation generalises to
\begin{align}\label{ODE_EGO-I}
    \Omega_\mathrm{DE}'(z) ={}& \left(1 - \Omega_\mathrm{DE}\right)\biggl[\frac{2\left(\alpha + \gamma H^{-4} - \mathcal{F}/H^2\right)}{\beta\,(1+z)}\nonumber\\
    &- \frac{3}{1+z} - \frac{\Omega_{r0}}{\Sigma}\biggr] ,
\end{align}
with $H(z)$ given by Eq.~\eqref{H_z}. The initial condition and numerical integration procedure are identical to the GO case; the GO limit is recovered for $\gamma = 0$.

\subsection{Cut-off 3}

Cut-off 3 replaces the $\gamma\,H^{-2}$ term with a term proportional to $\ddot{H}/H$:
\begin{equation}\label{EGO-II}
    L^{-2} = \alpha\, H^2 + \beta\,\dot{H} + \gamma\,\frac{\ddot{H}}{H}\,.
\end{equation}
The presence of $\ddot{H}$ introduces a genuine second-order time derivative which, upon conversion to redshift, promotes the evolution equation for $\Omega_\mathrm{DE}(z)$ to a second-order ODE. Imposing $L^{-2}=\mathcal{F}$ and using Eq.~\eqref{eq:Hdots}, one finds
\begin{equation}\label{ODE_EGO-II}
    \mathcal{E}' = \frac{2\left(\mathcal{F}/H^2-\alpha\right)}{\gamma\,(1+z)^2}
    + \frac{\beta-\gamma}{\gamma\,(1+z)}\,\mathcal{E} - \mathcal{E}^2 \ ,
\end{equation}
which, inserted in Eq.~\eqref{eq:ODE_EGOII}, yields the second-order equation for $\Omega_\mathrm{DE}(z)$; for $\Delta=0$ it reduces to the Bekenstein--Hawking case of Eq.~\eqref{eq:Eprime}.

Since the equation is second order, two boundary conditions are required. The first is the flatness constraint $\Omega_\mathrm{DE}(0) = 1 - \Omega_{m0} - \Omega_{r0}$. However, no second algebraic condition on $\Omega_\mathrm{DE}'(0)$ can be derived from the field equations; this quantity is therefore treated as a free shooting parameter and sampled alongside $\{\alpha, \beta, \gamma, C, \Delta\}$ in the MCMC analysis. Note that $\gamma$ multiplies the highest derivative: the GO limit $\gamma \to 0$ is singular and must be handled as a separate first-order system.

\section{Holographic Dark Energy Based on $S_3$ Generalized Entropy}\label{sec:S3}

We now extend the same construction to the three-parameter generalized entropy \cite{Nojiri:2022nof, Nojiri:2022aof, Nojiri:2022dkr}
\begin{equation}
    S_3 = \frac{1}{\gamma_s}\left[\left(1 + \frac{\alpha_s}{\beta_s} S\right)^{\beta_s} - 1\right] \ .
\end{equation}
This family of entropies was proposed as a unifying framework that encompasses, as special cases, the Tsallis--Cirto \cite{Tsallis:2012js}, R\'{e}nyi \cite{renyi1961measures}, Barrow \cite{Barrow:2020tzx}, Kaniadakis \cite{Kaniadakis:2005zk}, and Sharma--Mittal \cite{SayahianJahromi:2018irq} entropies, among others \cite{Nojiri:2022nof, Nojiri:2022aof}. Its application to HDE cosmology naturally interpolates between these distinct entropy-based scenarios and provides a flexible parametric test of the sensitivity of dark energy observables to the form of the underlying entropy functional. The four-parameter member of this family
has also been confronted with late-time BAO and supernova data~\cite{Odintsov:2025sew, Tyagi:2025zov}.

\subsection{GO cut-off}

Based on Eqs.~(\ref{rhode_Barrow}) and (\ref{fed}), applying the HDE construction with the GO cut-off, the fractional dark energy density takes the form
\begin{align}\label{S3ODE}
    \Omega_\mathrm{DE}(z) ={}& \frac{C^2}{\gamma_s H^2} \left(\alpha H^2 - (1+z)\beta H H'\right)^2 \nonumber\\
    &\times \left[\left(1 + \frac{\alpha_s}{\alpha \beta_s H^2 - (1+z)\beta \beta_s H H'}\right)^{\!\beta_s} - 1\right] \ .
\end{align}

For Eq.~(\ref{S3ODE}), not every parameter needs to have a specific value. We can reparameterize it to reduce the number of parameters. The reparameterized form is
\begin{align}\label{S3ODEr}
    \Omega_\mathrm{DE}(z) ={}& \frac{C_{r}^2}{H^2} \left(\alpha_{r} H^2 - (1+z)H H'\right)^2 \nonumber\\
    &\times \left[\left(1 + \frac{\alpha_{sr}}{\alpha_{r}H^2 - (1+z)H H'}\right)^{\!\beta_s} - 1\right] \ ,
\end{align}
where
\begin{align}
    C_{r}^2&=\frac{C^2 \beta^2}{\gamma_s} \ , \nonumber\\
    \alpha_{r}&=\frac{\alpha}{\beta}\ , \nonumber \\
    \alpha_{sr}&=\frac{\alpha_{s}}{\beta_{s}\beta} \ .
\end{align}
Unlike the Barrow case, the nonlinear and exponential dependence of $\Omega_\mathrm{DE}$ on $H'(z)$ prevents an explicit algebraic inversion for the derivative. Substituting into the Friedmann Eq.~\eqref{eq:Hubble} and differentiating with respect to $z$, one obtains a second-order ODE for $H(z)$ in which $H''(z)$ can be isolated as a function of the cosmological, entropic, and HDE parameters together with $H(z)$ and $H'(z)$. The explicit expression of $H''(z)$ is given in Appendix~\ref{app:H''}. Boundary conditions are imposed at $z = 0$: $H(0) = H_0$, while $H'(0)$ follows from Eq.~(\ref{eq:BC}).

\subsection{Cut-off 2}
Applying the HDE construction with cut-off 2 and performing reparameterization, the fractional dark energy density takes the form
\begin{align}\label{S3ODEIr}
    \Omega _{\mathrm{DE}}\left( z \right) &=\frac{C_{r}^{2}}{H^6}\left( \gamma _r+\alpha _rH^4-\left( 1+z \right) H^3H' \right) ^2 \nonumber
\\
&\times \left[ \left( \frac{\alpha _{sr}}{\frac{\gamma _r}{H^2}+\alpha _rH^2-\left( 1+z \right) HH'}+1 \right) ^{\beta _s}-1 \right] \ ,
\end{align}
where
\begin{align}
    C_{r}^2&=\frac{C^2 \beta^2}{\gamma_s} \ , \nonumber\\
    \alpha_{r}&=\frac{\alpha}{\beta}\ , \nonumber \\
    \gamma _r&=\frac{\gamma}{\beta}\ , \nonumber \\
    \alpha_{sr}&=\frac{\alpha_{s}}{\beta_{s}\beta} \ .
\end{align}
Following the same procedure, the explicit equation for $H''(z)$ is obtained (Appendix~\ref{app:H''}). Boundary conditions are imposed at $z = 0$: $H(0) = H_0$, while $H'(0)$ follows from Eq.~(\ref{eq:BC}).

\subsection{Cut-off 3}
Applying the HDE construction with cut-off 3 and performing reparameterization, the fractional dark energy density takes the form
\begin{align}\label{S3ODEIIr}
\Omega _{\mathrm{DE}}\left( z \right) ={}&\frac{C_{r}^{2}}{H^4}A_1^2 \nonumber
\\
&\times \left[ \left( \frac{\alpha _{sr}H}{A_1}+1 \right) ^{\beta _s}-1 \right]\ ,
\end{align}
where
\begin{align}
           A_1&=\alpha _rH^3-\left( 1+z \right) H^2H'+\gamma _r\left( 1+z \right) H^2H' \nonumber\\
           &+\gamma _r\left( 1+z \right) ^2{HH^{'}}^2 +\gamma _r\left( 1+z \right) ^2H^2H'' \ , \nonumber\\
    C_{r}^2&=\frac{C^2 \beta^2}{\gamma_s} \ , \nonumber\\
    \alpha_{r}&=\frac{\alpha}{\beta}\ , \nonumber \\
    \gamma _r&=\frac{\gamma}{\beta}\ , \nonumber \\
    \alpha_{sr}&=\frac{\alpha_{s}}{\beta_{s}\beta} \ .
\end{align}
Unlike the GO cut-off and cut-off 2, $\Omega_\mathrm{DE}$ now depends on $H''(z)$, so that the closure relation obtained by substituting Eq.~(\ref{S3ODEIIr}) into Eq.~(\ref{eq:Hubble}) and differentiating with respect to $z$ determines the third derivative $H'''(z)$ (Appendix~\ref{app:H''}). Boundary conditions are imposed at $z = 0$: $H(0) = H_0$, $H'(0)=H'_0$ and $H''(0)=H''_0$. Either $H'_0$ or $H''_0$ can be chosen as a free parameter, the other then being fixed by Eq.~(\ref{eq:BC}). We choose $H''_0$ as the free parameter and solve for $H'_0$: this is numerically much cheaper and keeps $H'_0$ close to its $\Lambda$CDM value.

\section{Holographic Dark Energy Based on $S_5$ Generalized Entropy}\label{sec:S5}
Although the three- and four-parameter entropies~\cite{Nojiri:2022nof,Nojiri:2022aof} reduce to other known entropies, they become singular in some cosmological scenarios. In particular, in bounce cosmology they diverge at the instant of the bounce, when the Hubble parameter vanishes. This motivates a generalized entropy that reproduces the known entropy functions while remaining non-singular during the entire cosmological evolution. Such a singularity-free five-parameter entropy was proposed in \cite{Odintsov:2022qnn}, and has the form
\begin{align}
		S_5={}&\frac{1}{\gamma_s}\left\{ \left[ 1+\frac{1}{\varepsilon_s}\tanh  \left( \frac{\varepsilon_s \alpha _{s+}}{\beta_s}S \right) \right] ^{\beta_s}\right.\nonumber
        \\
        &\left. -\left[ 1+\frac{1}{\varepsilon_s}\tanh \left( \frac{\varepsilon_s \alpha _{s-}}{\beta_s}S \right) \right]^{-\beta_s} \right\},  \label{S5tanh}
\end{align}
where $\alpha _{s\pm}$, $\beta_s$, $\gamma_s$ and $\varepsilon_s$ are the parameters which are considered to be positive, and $S$ denotes the Bekenstein--Hawking entropy. Owing to the $\tanh$ function, this entropy remains finite, and thus non-singular, during the whole cosmological evolution, and for suitable choices of the parameters it reduces to various known entropies, namely the Bekenstein--Hawking, Tsallis, Barrow, R\'enyi, Kaniadakis and Sharma--Mittal entropies, as well as the entropy arising in Loop Quantum Gravity.

The five-parameter entropy admits more than one representation. In the present paper we adopt the form
\begin{align}
	S_5={}&\frac{1}{\gamma_s}\left\{ \left[ 1+\frac{1}{\varepsilon_s}\arctan \left( \frac{\varepsilon_s \alpha _{s+}}{\beta_s}S \right) \right] ^{\beta_s}\right.\nonumber
    \\
    &\left. -\left[ 1+\frac{1}{\varepsilon_s}\arctan \left( \frac{\varepsilon_s \alpha _{s-}}{\beta_s}S \right) \right]^{-\beta_s} \right\} \ .  \label{S5arc}
\end{align}
which is obtained from Eq.~(\ref{S5tanh}) by replacing $\tanh$ with $\arctan$. Eq.~(\ref{S5arc}) is still finite and reduces to the same known entropies for suitable choices of the parameters.

\subsection{GO cut-off}
Applying the HDE construction with the GO cut-off and performing reparameterization, the fractional dark energy density takes the form
\begin{align}\label{S5ODEr}
\Omega_\mathrm{DE}(z) ={}& \frac{C_{r}^2}{H^2} \left(\alpha_{r} H^2 - (1+z)H H'\right)^2 \nonumber\\
    &\times \left\{ \left[ 1+\frac{1}{\varepsilon_s}\arctan \frac{\alpha_{sr1}}{\alpha_{r}H^2 - (1+z)H H'} \right] ^{\beta_s}\right.\nonumber
    \\
    &\left. -\left[ 1+\frac{1}{\varepsilon_s}\arctan \frac{\alpha_{sr2}}{\alpha_{r}H^2 - (1+z)H H'} \right]^{-\beta_s} \right\} \ ,
\end{align}
where
\begin{align}
    C_{r}^2&=\frac{C^2 \beta^2}{\gamma_s} \ , \nonumber\\
    \alpha_{r}&=\frac{\alpha}{\beta}\ , \nonumber \\
    \alpha_{sr1}&=\frac{\alpha_{s+}\varepsilon_s}{\beta_{s}\beta} \ , \nonumber\\
    \alpha_{sr2}&=\frac{\alpha_{s-}\varepsilon_s}{\beta_{s}\beta} \ .
\end{align}
As in the previous section, the explicit expression of $H''(z)$ can be obtained (Appendix~\ref{app:H''}). Boundary conditions are imposed at $z = 0$: $H(0) = H_0$, while $H'(0)$ follows from Eq.~(\ref{eq:BC}).

\subsection{Cut-off 2}
Applying the HDE construction with cut-off 2 and performing reparameterization, the fractional dark energy density takes the form
\begin{align}\label{S5ODEIr}
    &\Omega _{\mathrm{DE}}\left( z \right) ={}\frac{C_{r}^{2}}{H^2}\left( \frac{\gamma _r}{H^2}+\alpha _rH^2-\left( 1+z \right) HH' \right) ^2 \nonumber
\\
&\times \left\{ \left[ 1+\frac{1}{\varepsilon_s}\arctan \left( \frac{\alpha_{sr1}}{\frac{\gamma _r}{H^2}+\alpha _rH^2-\left( 1+z \right) HH'} \right) \right] ^{\beta_s}\right.\nonumber
    \\
    &\left. -\left[ 1+\frac{1}{\varepsilon_s}\arctan \left( \frac{\alpha_{sr2}}{\frac{\gamma _r}{H^2}+\alpha _rH^2-\left( 1+z \right) HH'} \right) \right]^{-\beta_s} \right\} \ ,
\end{align}
where
\begin{align}
    C_{r}^2&=\frac{C^2 \beta^2}{\gamma_s} \ , \nonumber\\
    \alpha_{r}&=\frac{\alpha}{\beta}\ , \nonumber \\
    \gamma _r&=\frac{\gamma}{\beta}\ , \nonumber \\
    \alpha_{sr1}&=\frac{\alpha_{s+}\varepsilon_s}{\beta_{s}\beta} \ , \nonumber\\
    \alpha_{sr2}&=\frac{\alpha_{s-}\varepsilon_s}{\beta_{s}\beta} \ .
\end{align}
Following the same procedure, the explicit equation for $H''(z)$ is obtained (Appendix~\ref{app:H''}). Boundary conditions are imposed at $z = 0$: $H(0) = H_0$, while $H'(0)$ follows from Eq.~(\ref{eq:BC}).

\subsection{Cut-off 3}
Applying the HDE construction with cut-off 3 and performing reparameterization, the fractional dark energy density takes the form
\begin{align}\label{S5ODEIIr}
\Omega _{\mathrm{DE}}\left( z \right) ={}&\frac{C_{r}^{2}}{H^4}A_2^2 \nonumber
\\
&\times \left\{ \left[ 1+\frac{1}{\varepsilon_s}\arctan \left( \frac{\alpha_{sr1}H}{A_2} \right) \right] ^{\beta_s}\right.\nonumber
    \\
    &\left. -\left[ 1+\frac{1}{\varepsilon_s}\arctan \left( \frac{\alpha_{sr2}H}{A_2} \right) \right]^{-\beta_s} \right\}\ ,
\end{align}
where
\begin{align}
A_2&=\alpha _rH^3-\left( 1+z \right) H^2H'+\gamma _r\left( 1+z \right) H^2H' \nonumber\\
&+\gamma _r\left( 1+z \right) ^2{HH^{'}}^2 +\gamma _r\left( 1+z \right) ^2H^2H'' \ , \nonumber\\
C_{r}^2&=\frac{C^2 \beta^2}{\gamma_{s}} \ , \nonumber\\
\alpha_{r}&=\frac{\alpha}{\beta}\ , \nonumber \\
\gamma_r&=\frac{\gamma}{\beta}\ , \nonumber \\
\alpha_{sr1}&=\frac{\alpha_{s+}\varepsilon_{s}}{\beta_{s}\beta} \ , \nonumber\\
\alpha_{sr2}&=\frac{\alpha_{s-}\varepsilon_{s}}{\beta_{s}\beta} \ .
\end{align}
As for the $S_3$ entropy, substituting Eq.~(\ref{S5ODEIIr}) into Eq.~(\ref{eq:Hubble}) and differentiating with respect to $z$ yields an explicit equation for $H'''(z)$ (Appendix~\ref{app:H''}). Boundary conditions are imposed at $z = 0$: $H(0) = H_0$, $H'(0)=H'_0$ and $H''(0)=H''_0$; we choose $H''_0$ as the free parameter and solve for $H'_0$.

\section{Observational datasets}\label{sec:obs_data}
In this section, we describe the various observational datasets we used to constrain cosmological parameters and their joint analysis. Specifically, we consider the following datasets: Baryon Acoustic Oscillations (BAO), Type Ia Supernovae (SNe Ia), and Cosmic Microwave Background (CMB). 

\begin{itemize}
    \item \textbf{Type Ia Supernovae (SNe Ia)}: \\
    We consider the \textit{Pantheon$+$} dataset\footnote{The repository of the Pantheon$+$ sample can be found at the following link: \url{https://github.com/PantheonPlusSH0ES/DataRelease} .} \cite{Brout_2022}, labeled hereinafter as Pan+, which consist of 1550 data points in the redshift range $0.001\leq z\leq 2.26$.
    Each data point is given by the observed distance modulus at a given redshift $\mu^\text{obs}(z)$. The $\chi^2$ value can be computed as
    \begin{equation}
        \chi^2_{\text{SNeIa}} = \sum_{i,j=1}^{1550}\Delta\mu_i\Big(\text{Cov}^{-1}\Big)_{ij}\Delta\mu_j \ ,
    \end{equation}
 where $\Delta\mu_i = (\mu^\text{obs}(z)-\mu^\text{th}(z))_i$ and $\text{Cov}$ is the covariant matrix of the dataset obtained summing the statistical and systematic covariance matrices. The theoretical expression of the distance modulus is 
    \begin{equation}
    \mu^\text{th}(z) = 5 \, \log_{10}\left(\frac{D_L(z)}{10 \text{pc}}\right) \ ,
    \end{equation}
    which is related to the luminosity distance $D_L(z)$:
    \begin{equation}
        D_\mathrm{L}(z) = (1+z) \int_0^z\frac{dz'}{H(z')} \ .
    \end{equation}
    This latter quantity depends on the considered cosmological model and encodes information about the cosmological dynamics through the presence of the Hubble parameter.

    \item \textbf{Baryon Acoustic Oscillations (BAO):} \\
    We use of DESI DR2 measurements of BAO \cite{DESI:2025zgx} obtained from galaxy and quasar data. Specifically, we considered the measurements of $D_M(z) / r_d$ and $D_H(z)/r_d$ for six redshift bins from $0.1\leq z\leq4.2$, where $D_H(z) \equiv 1/H(z)$ is the Hubble distance in natural units ($c=1$), $D_M(z)$ is the transverse comoving distance, related to the luminosity distance through the relation $D_L(z)=(1+z)D_M(z)$, and $r_d$ is the sound horizon at the baryon drag epoch. The value of $r_d$ can be derived from its definition through an integral over the redshift, but following the approach of the DESI collaboration \cite{DESI:2025zgx}, we use the fitting formula \cite{Brieden_2023}
    \begin{align}  
        r_d ~=~ & 147.05\,{\rm Mpc} \,\nonumber\times\\
        &  \left(\frac{\omegab}{0.02236}\right)^{-0.13} 
               \left(\frac{\omegabc}{0.1432}\right)^{-0.23}    
            \left(\frac{\Neff}{3.04}\right)^{-0.1}\ ,
        \label{eq:rdformula}
    \end{align}
    where the standard notation is used, $\omegab = \Omegab h^2$ and $\omegabc = (\Omegab+\Omegac) h^2$.
    We consider the number of effective neutrino species to be constant $\Neff=3.04$, since varying this parameter does not produce relevant changes in dark energy analysis \cite{DESI:2025zgx}.

    \item \textbf{Cosmic Microwave Background (CMB):}\\
    We use the compression of the full information of the Planck 2018 data \cite{Planck_2020} on the parameters $(\theta_\ast,\omegab,\omega_{cb})$, based on the \texttt{CamSpec} CMB likelihood as computed in \cite{Lemos:2023xhs}. We assume $T_\text{CMB} = 2.7255$ K \cite{Fixsen:2009ug}. We use this analysis since these quantities are related to the early epoch of the Universe, in which the Holographic and entropic models replicates the $\Lambda$CDM case. The value of the acoustic angular scale is defined as $\theta_\ast = r_\ast / D_M(z_\ast)$, where $z_\ast$ is the redshift at the end of recombination and $r_\ast$ is the sound horizon at this redshift. \\
    The dataset is given by the mean vector
    \begin{equation}
        \mu^\text{obs} = \begin{pmatrix}
        \theta_\ast\\
        \omegab\\
        \omegabc
        \end{pmatrix} = 
        \begin{pmatrix}
        0.01041 \\
        0.02223 \\
        0.14208
        \end{pmatrix}
        \label{eq:CMB_compression_mean}
    \end{equation}
    and covariance
    \begin{equation}
        \text{Cov} = 10^{-9} \times 
        \begin{pmatrix}
        0.006621 &  0.12444 & -1.1929 \\
        0.12444 &  21.344 & -94.001 \\
        -1.1929 & -94.001 & 1488.4
        \end{pmatrix}\,.
        \label{eq:CMB_compression_cov}
    \end{equation}
    To perform a fully rigorous analysis of the CMB dataset, one should derive the parameters $z_\ast$ and $r_\ast$ through a prior analysis. We keep fixed $z_\ast = 1089$, and we used the formula\footnote{This formula has been computed for the case of fixed neutrino mass, but it can be shown that a variation of the latter does not produce relevant changes in the constant of proportionality.} for $r_\ast = r_d/1.01841$ from \cite{Hu_1996}. \\
    We stress that this compressed CMB likelihood is based on the acoustic scale $\theta_*$ and the physical baryon and cold dark matter densities, which are primarily sensitive to the background expansion history. 
\end{itemize}

We perform a standard Bayesian parameter inference \cite{padilla2021cosmological} using a Markov Chain Monte Carlo (MCMC) approach to obtain the posterior distributions of the model parameters. The analysis is carried out with the publicly available \texttt{emcee} Python package \cite{Foreman-Mackey2019}.

For each MCMC sample, the evolution equations of Secs.~\ref{sec:BH}--\ref{sec:S5} (for $\Omega_\mathrm{DE}(z)$ in the Barrow sector, for $H(z)$ in the $S_3$ and $S_5$ sectors) are integrated numerically from $z=0$ to $z \simeq 1100$ using \texttt{solve\_ivp}, with the flatness condition $\Omega_\mathrm{DE}(0) = 1 - \Omega_{m,0} - \Omega_{r,0}$. The reconstructed $H(z)$ is then used to compute the observables entering the likelihood. For cut-off 3, whose equations are of higher order for all three entropies, the additional boundary value ($\Omega_\mathrm{DE}'(0)$ for Barrow, $H''_0$ for $S_3$ and $S_5$) cannot be derived algebraically and is treated as a free shooting parameter sampled alongside the physical parameters. The singular limit $\gamma \to 0$, in which the equation reduces to first order, is handled as a separate branch in the integrator.

The present-day radiation density is fixed to $\Omega_{r,0} = 2.469 \times 10^{-5}\, h^{-2}$ with an effective number of neutrino species $N_\mathrm{eff} = 3.044$; as shown in \cite{DESI:2025zgx}, this assumption has negligible impact on the results. The absolute magnitude $M_b$ and the baryon density $\omega_b \equiv \Omega_{b,0}\,h^2$ are sampled as nuisance parameters entering the Pantheon+ and CMB likelihoods, respectively.

The parameter space and uniform priors adopted for each entropy--cut-off combination are summarized in Table~\ref{tab:priors}. Common to all models are the base parameters $\{H_0,\,\Omega_{m,0},\,C,\,M_b,\,\omega_b\}$, supplemented by the cut-off parameters $\{\alpha,\beta\}$ (and $\gamma$ for cut-offs 2 and 3) and the entropy-specific parameters: $\Delta$ for Barrow, $\{\alpha_{sr},\beta_s\}$ for $S_3$, and $\{\alpha_{sr1},\alpha_{sr2},\beta_s,\varepsilon_s\}$ for $S_5$. For $S_3$ and $S_5$ the reparameterized quantities $C_r$, $\alpha_r$, $\gamma_r$ of Secs.~\ref{sec:S3} and~\ref{sec:S5} are sampled in place of $C$, $\alpha$, $\beta$, $\gamma$.

\begin{table}[t]
\centering
\caption{Uniform priors on the sampled parameters. The full parameter space for a given model is the union of the base parameters, the cutoff-specific block, and the entropy-specific block. For cut-off 3 models an additional shooting parameter is required (bottom block). The Hubble constant is in units of $\mathrm{km\,s^{-1}\,Mpc^{-1}}$.}
\label{tab:priors}
\renewcommand{\arraystretch}{1.25}
\centering
{%
\begin{tabular}{llc}
\hline\hline
\multicolumn{3}{c}{\textit{Base parameters (all models)}} \\[2pt]
$H_0$            & $[40,\,100]$    & \\
$\Omega_{m0}$   & $[0.01,\,0.5]$  & \\
$C/C_r$              & $[0.01,\,30]$   & \\
$M_b$            & $[-20,\,-19]$   & \\
$\omega_b$       & $[0.01,\,0.1]$  & \\[4pt]
\hline
\multicolumn{3}{c}{\textit{Cut-off parameters (Barrow)}} \\[2pt]
Parameter        & Prior           & Cut-off \\[2pt]
$\alpha$         & $[0.01,\,10]$   & GO, 2, 3 \\
$\beta$          & $[0.005,\,5]$   & GO, 2, 3 \\
$\gamma$         & $[0.01,\,5]$    & 2, 3 \\[4pt]
\hline
\multicolumn{3}{c}{\textit{Cut-off parameters ($S_3$ and $S_5$)}} \\[2pt]
Parameter        & Prior           & Cut-off \\[2pt]
$\alpha_r$         & $[-30,\,30]$   & GO, 2, 3 \\
\multirow{2}{*}{$\gamma_r$}         & $[-30,\,30]$    & 3 \\
              &$[-150,\,150]$    & 2 \\[4pt]
\hline
\multicolumn{3}{c}{\textit{Entropy parameters}} \\[2pt]
Parameter        & Prior           & Entropy \\[2pt]
$\Delta$         & $[0,\,1]$       & Barrow \\
$\alpha_{sr}$ & $[-30,\,30]$ & $S_3$ \\
$\beta_{s}$ & $[0,\,30]$ & $S_3$, $S_5$ \\
$\alpha_{sr1},\;\alpha_{sr2}$   & $[-70,\,70]$  & $S_5$ \\
$\varepsilon_s$     & $[0,\,100]$     & $S_5$ \\[4pt]
\hline
\multicolumn{3}{c}{\textit{Shooting parameter (cut-off 3 only)}} \\[2pt]
$\Omega_\mathrm{DE}'(0)$ & $[-5,\,5]$ & Barrow \\
$H_0''$          & $[-200,\,200]$  & $S_3$, $S_5$ \\[2pt]
\hline\hline
\end{tabular}%
}
\end{table}

\section{Results}\label{sec:results}

We now present the observational constraints on the nine entropy--cut-off combinations obtained from the joint DESI DR2 BAO\,+\,Pantheon+\,+\,CMB dataset. The full posterior distributions are collected in Appendix~\ref{app:corner}; here we summarize the main findings, organized by entropy sector. Tables~\ref{tab:barrow_results},~\ref{tab:S3_results}, and~\ref{tab:S5_results} list the marginalised constraints (68\% C.L.) and the information criteria $\Delta\mathrm{AIC}$ and $\Delta\mathrm{BIC}$ \cite{Akaike:1974,Schwarz:1978,Liddle:2007fy}. Throughout, $\Delta\mathrm{AIC}$ and $\Delta\mathrm{BIC}$ are computed with respect to the overall best-fitting model, Barrow\,+\,cut-off 3, so that the three tables can be compared directly.

\subsection{Baseline cosmological parameters}

Across all nine models the Hubble constant remains stable, $H_0 \simeq 67.0$--$68.1\;\mathrm{km\,s^{-1}\,Mpc^{-1}}$, and the nuisance parameters $M_b$ and $\omega_b$ are insensitive to the choice of entropy and cut-off. The matter density falls into two groups. The three Barrow models, $S_3$\,+\,GO and $S_5$\,+\,cut-off 3 give $\Omega_{m,0} \simeq 0.310$--$0.316$, while the remaining four models ($S_3$ with cut-offs 2 and 3, $S_5$ with the GO cut-off and cut-off 2) settle at $\Omega_{m,0} \simeq 0.283$--$0.284$. The latter group coincides with the models carrying the largest information-criteria penalties. The largest shift in $H_0$ occurs for $S_5$\,+\,cut-off 3, which gives $H_0 = 67.03^{+0.69}_{-0.51}$ together with $\Omega_{m,0} = 0.316^{+0.006}_{-0.005}$.

\subsection{Barrow entropy}\label{sec:res_barrow}

The Barrow results are summarized in Table~\ref{tab:barrow_results} and Figs.~\ref{fig:corner_Barrow_GO}--\ref{fig:corner_Barrow_EGOII}.

With the GO cut-off, the Barrow exponent is tightly bounded to $\Delta < 0.00432$ (68\% C.L.), consistent with the Bekenstein--Hawking limit $\Delta = 0$. The GO parameters $\alpha$ and $\beta$ are well determined and display a pronounced positive degeneracy. The holographic parameter takes a moderate value, $C = 0.543^{+0.057}_{-0.048}$, and the model is strongly disfavoured relative to cut-off 3 ($\Delta\mathrm{AIC} = 148.07$, $\Delta\mathrm{BIC} = 137.31$).

Moving to cut-off 2 leaves this picture essentially unchanged: the additional parameter $\gamma$ is entirely prior-dominated ($\gamma > 6.88$ at 68\% C.L.; Fig.~\ref{fig:corner_Barrow_EGOI}), confirming that the $\gamma\,H^{-2}$ correction is irrelevant within the redshift range probed by the data. All remaining parameters, including $\Delta < 0.00433$, are consistent with the GO values, and the information criteria worsen by roughly the penalty of the extra parameter ($\Delta\mathrm{AIC} = 150.59$).

Cut-off 3 produces a qualitatively different result. The $\ddot{H}/H$ term introduces a genuinely second-order ODE whose shooting parameter $\Omega'_\mathrm{DE}(0)$ is tightly constrained to $-0.594^{+0.017}_{-0.016}$ (Fig.~\ref{fig:corner_Barrow_EGOII}), demonstrating that the data are sensitive to the second-order dynamics. The cut-off parameter $\gamma = 1.21^{+0.21}_{-0.35}$ is clearly non-zero. This richer structure allows the Barrow exponent to move away from zero, $\Delta = 0.28^{+0.28}_{-0.20}$, which is nevertheless compatible with $\Delta=0$ at about $1.4\sigma$, accompanied by a large holographic parameter $C = 13.4^{+4.2}_{-6.1}$ that absorbs the resulting $C$--$\Delta$ degeneracy. This model achieves the best fit of the whole analysis, the improvement in $\chi^2$ largely outweighing the penalty for its two additional parameters.

\subsection{$S_3$ entropy}\label{sec:res_S3}

The $S_3$ results are given in Table~\ref{tab:S3_results} and Figs.~\ref{fig:corner_S3_GO}--\ref{fig:corner_S3_EGOII}.

The GO cut-off yields the best fit in this sector and the second-best overall ($\Delta\mathrm{AIC} = 3.09$). Both entropy parameters are well constrained ($\alpha_{sr} = 12.987^{+2.130}_{-2.086}$, $\beta_s = 1.212^{+0.113}_{-0.075}$), and the cut-off parameter $\alpha_r = 1.797^{+0.190}_{-0.110}$ is determined with $\sim$10\% precision. The larger $\Delta\mathrm{BIC} = 24.61$ reflects the BIC penalty for the larger number of parameters relative to Barrow\,+\,cut-off 3.

Cut-off 2 fails to improve the fit: $\gamma_r = 2.626^{+81.647}_{-78.472}$, which means the data provides almost no constraint on it (Fig.~\ref{fig:corner_S3_EGOI}), and the information criteria degrade sharply ($\Delta\mathrm{AIC} = 142.98$). The cut-off parameter collapses to a narrow peak at $\alpha_r \simeq 1.640$ with sub-percent uncertainties, while the entropy parameters $\alpha_{sr}$ and $\beta_s$ shift substantially from their GO values. This reorganisation of the parameter space reflects the degeneracy introduced by $\gamma_r$: the sampler settles on an effectively lower-dimensional surface at fixed $\alpha_r$ where the $\gamma_r$-dependent terms are negligible.

Cut-off 3 similarly provides no gain over GO ($\Delta\mathrm{AIC} = 144.73$). The shooting parameter $H_0''$ is prior-dominated (Fig.~\ref{fig:corner_S3_EGOII}), indicating that the $S_3$ functional does not generate sufficient structure in the second-order dynamics to be constrained by the data. The cut-off parameter $\gamma_r$ is determined but consistent with zero ($\gamma_r = -0.002^{+0.002}_{-0.003}$), reinforcing the GO limit.

\subsection{$S_5$ entropy}\label{sec:res_S5}

The $S_5$ results appear in Table~\ref{tab:S5_results} and Figs.~\ref{fig:corner_S5_GO}--\ref{fig:corner_S5_EGOII}.

With the GO cut-off, the four entropy parameters $\{\alpha_{sr1}, \alpha_{sr2}, \beta_s, \varepsilon_s\}$ display broad, mutually correlated posteriors (Fig.~\ref{fig:corner_S5_GO}), indicating that the five-parameter entropy functional is over-parameterized for the current dataset. The cut-off parameter again locks onto $\alpha_r \simeq 1.640$, and the model is heavily penalized ($\Delta\mathrm{AIC} = 145.02$, $\Delta\mathrm{BIC} = 177.31$).

Cut-off 2 confirms the general pattern: $\gamma_r$ is unconstrained ($\gamma_r = -10.099^{+46.796}_{-35.573}$; Fig.~\ref{fig:corner_S5_EGOI}), the entropy parameters remain poorly determined, and the information criteria worsen further ($\Delta\mathrm{AIC} = 146.63$, $\Delta\mathrm{BIC} = 184.30$).

$S_5$\,+\,cut-off 3 is the only model in this sector for which the extended cut-off makes a material difference. The shooting parameter $H_0''$ is well determined ($24.42^{+2.86}_{-2.32}$; Fig.~\ref{fig:corner_S5_EGOII}), and both cut-off parameters are tightly constrained, $\alpha_r = 0.344^{+0.035}_{-0.043}$ and $\gamma_r = 0.253^{+0.009}_{-0.008}$, with $\gamma_r$ far from zero. The entropy parameters also tighten considerably compared to GO. The fit improves dramatically with respect to the other $S_5$ models ($\Delta\mathrm{AIC} = 17.61$ versus $\simeq 145$), at the cost of a shift in the base cosmological parameters ($H_0 \simeq 67.46$, $\Omega_{m,0} \simeq 0.312$); the large parameter count keeps it well behind Barrow\,+\,cut-off 3 in BIC ($\Delta\mathrm{BIC} = 61.42$).

\subsection{Cross-cutting patterns}\label{sec:res_patterns}

Several results are robust across all entropy sectors and deserve separate emphasis.

Cut-off 2 ($\gamma\,H^{-2}$) never improves the fit. In all three entropy cases the posterior on $\gamma$ ($\gamma_r$) is prior-dominated, in line with the expectation that this term only becomes dynamically relevant in the asymptotic future $H \to 0$, well outside the sensitivity of current cosmological probes. With respect to cut-off 2, the GO cut-off is therefore the data-preferred truncation.

Cut-off 3 ($\gamma\,\ddot{H}/H$) exhibits a sharp dichotomy. For Barrow and $S_5$, the second-order dynamics is genuinely probed by the data, as evidenced by the tight constraints on $\Omega'_\mathrm{DE}(0)$ or $H_0''$ and on $\gamma$, and the fit improves with respect to GO by $\Delta\mathrm{AIC} \simeq 148$ and $\simeq 127$, respectively. For $S_3$, the shooting parameter remains unconstrained and $\gamma_r$ is driven to zero, suggesting that the second-order correction cannot be distinguished from the GO baseline in this sector.

Finally, model comparison across all nine models reveals a clear hierarchy. Barrow\,+\,cut-off 3 provides the best fit, followed by $S_3$\,+\,GO ($\Delta\mathrm{AIC} = 3.09$, $\Delta\mathrm{BIC} = 24.61$) and $S_5$\,+\,cut-off 3 ($\Delta\mathrm{AIC} = 17.61$); all other models are disfavoured by $\Delta\mathrm{AIC} \gtrsim 140$. In particular, Barrow\,+\,GO, which effectively reduces to the Bekenstein--Hawking case since $\Delta < 0.0043$, is among the disfavoured models. At the same time, even in the best-fitting model the Barrow exponent remains compatible with zero at about $1.4\sigma$, so the preference expressed by the data concerns the structure of the cut-off rather than a departure from the area law.

\subsection{The evolution of dark energy}\label{sec:evo_dark}

In particular, we also investigated the evolution of dark energy for $S_3$ and $S_5$ under these three types of cut-offs; see Figs.~\ref{fig:S3_dark}--\ref{fig:S5_dark}.

Regarding $S_3$ entropy, $S_3$\,+\,cut-off 2 and $S_3$~+~cut-off 3 exhibit similar evolutionary behavior of dark energy. $S_3$\,+\,GO exhibits quintessence-like behavior, while $S_3$\,+\,cut-off 2 and $S_3$\,+\,cut-off 3 demonstrate a quintessence-to-phantom transition. These behaviors in
contrast to the recent claims of dynamical dark energy at higher redshifts\cite{DESI:2024mwx}, where a late-time quintessence with phantom-crossing was observed using the standard Chavelier-Polarski-Linder (CPL) parameterization \cite{Chevallier:2000qy, Linder:2002et}.

Regarding $S_3$ entropy, $S_5$\,+\,GO and $S_5$~+~cut-off 2 exhibit similar evolutionary behavior of dark energy. $S_5$\,+\,GO and $S_5$\,+\,cut-off 2 demonstrate a quintessence-to-phantom transition, while $S_5$\,+\,cut-off 3 exhibits a mild phantom-to-quintessence transition. The value of $\omega_\mathrm{DE}(z)$ of $S_5$\,+\,cut-off 3 consistently remains near -1 throughout its evolution, which implies that the dark energy density $\rho_\mathrm{DE}$ is very close to a constant, but with a slight deviation.

\section{Conclusions}
\label{sec:conclusions}

We have carried out a systematic feasibility analysis of holographic dark energy
built from generalized entropy functionals and extended infrared cut-offs,
constrained by a joint DESI DR2 BAO + Pantheon+ + CMB dataset. The study spans a
$3\times 3$ grid of nine models---Barrow, $S_3$ and $S_5$ entropies, each combined
with the GO cut-off and cut-offs 2 and 3---with the Bekenstein--Hawking case recovered
in the $\Delta\to 0$ limit of the Barrow models. Our main findings are as follows.

\emph{The background is well anchored.} Across all nine models the Hubble constant
remains in the range $H_0\simeq 67.0$--$68.1\,\mathrm{km\,s^{-1}\,Mpc^{-1}}$ and the nuisance parameters $M_b$ and $\omega_b$
are insensitive to the entropy and cut-off. The matter density takes values
$\Omega_{m,0}\simeq 0.310$--$0.312$ in the well-fitting models and
$\Omega_{m,0}\simeq 0.284$ in the four $S_3$ and $S_5$ models with the largest
information-criteria penalties.

\emph{Cut-off 2 is unconstrained.} In all three entropy
sectors the parameter multiplying the $H^{-2}$ term is entirely
prior-dominated. This is a physically transparent result: the $\gamma H^{-2}$ term becomes dynamically relevant
only in the asymptotic future ($H\to 0$), well outside the redshift range probed by
current data, and it never improves the fit with respect to the GO cut-off.

\emph{Cut-off 3 shows a sharp dichotomy.} For the Barrow and $S_5$
entropies the second-order dynamics is genuinely probed: the shooting parameter
$\Omega'_{\rm DE}(0)$ (Barrow) or $H''_0$ ($S_5$) and the parameter $\gamma$ are tightly
constrained, and the fit improves with respect to GO by $\Delta\mathrm{AIC}\simeq 148$
(Barrow) and $\simeq 127$ ($S_5$). For $S_3$, by contrast, the shooting parameter
remains prior-dominated and $\gamma_r$ is driven to zero.

\emph{Model comparison.} The overall best fit is Barrow~+~cut-off 3, followed by
$S_3$~+~GO ($\Delta\mathrm{AIC}=3.09$) and $S_5$~+~cut-off 3 ($\Delta\mathrm{AIC}=17.61$),
while all remaining models, including Barrow~+~GO, are disfavoured by
$\Delta\mathrm{AIC}\gtrsim 140$. The BIC, which penalizes the larger parameter count of the
multi-parameter entropies more strongly, confirms Barrow~+~cut-off 3 as the preferred model
($\Delta\mathrm{BIC}=24.61$ for $S_3$~+~GO and $60.66$ for $S_5$~+~cut-off 3).

\emph{The Barrow exponent stays near the classical limit.} With the GO cut-off and
cut-off 2, $\Delta$ is tightly bounded, $\Delta\lesssim 0.004$ (68\% C.L.),
consistent with the Bekenstein--Hawking value $\Delta=0$. Only Barrow~+~cut-off 3
allows a non-zero central value, $\Delta=0.28^{+0.28}_{-0.20}$, which is still
compatible with zero at about $1.4\sigma$ and is accompanied by a large holographic
parameter ($C=13.4^{+4.2}_{-6.1}$) that absorbs a pronounced $C$--$\Delta$ degeneracy. Present data
therefore provide no compelling evidence for a fractal deformation of the horizon.

\emph{The multi-parameter entropies are over-parameterized.} The $S_3$ and,
especially, $S_5$ functionals display broad and strongly correlated posteriors: the
data effectively select a lower-dimensional subspace, leaving much of the entropic
freedom inert. Tightening these constraints will require either datasets beyond DESI
DR2 or physically motivated priors on the entropy parameters, for instance from
primordial gravitational waves, which already restrict the viable entropic parameter
space~\cite{Odintsov:2024sbo,DOnofrio:2025bol}.

\emph{Phantom crossing.} Among the six models of $S_3$ and $S_5$ entropies, five exhibit phantom crossing or inverse phantom crossing. $S_3$\,+\,cut-off 2 and $S_3$\,+\,cut-off 3, in particular, stand in contrast to the CPL parameterization with DESI data; this suggests that our models do not rely on the CPL parameterization and indicates that the phantom-to-quintessence transition is merely an artifact of a specific parameterization.

Taken together, these results paint a coherent picture. Meaningful extensions of the
GO cut-off must act at the intermediate redshifts where BAO and supernovae carry
constraining power (as cut-off 3 does in the Barrow and $S_5$ sectors), rather than in
asymptotic regimes (as cut-off 2 does). The data favour such a higher-order cut-off
in two of the three entropy sectors, while showing no significant preference for a
departure from the Bekenstein--Hawking area law in the Barrow sector: the extra
freedom required by the data lies in the infrared cut-off rather than in the entropy.

Several extensions are natural. A split-dataset analysis of the best-fitting Barrow~+~cut-off 3
model would clarify which probes drive the constraints on $C$, $\gamma$ and $\Delta$ and how
their degeneracy is broken. Adding growth-of-structure observables (redshift-space
distortions, weak lensing) would test the perturbation sector and could lift
degeneracies invisible to background data alone. Finally, alternative higher-order
cut-off structures, such as an $H^4/\dot H$ proposal, can be accommodated within
the same MCMC framework and are left for future work. As DESI, Euclid and LSST
sharpen late-time constraints, the systematic baseline established here provides a
reference against which more refined tests of holographic dark energy can be
measured.

\section*{Acknowledgments}
KC acknowledges support from National Scholarship Council (PhD fellowship from the China Scholarship Council (CSC) (No. 202506180029)). SD is funded by MCIN/AEI/10.13039/501100011033 and FSE+, reference PRE2021-098098 and the program Unidad de Excelencia María de Maeztu
CEX2020-001058-M. TP is funded by Advanced Research Grant-MATRICS (ARG-MATRICS) of
Anusandhan National Research Foundation (ANRF): ANRF/ARGM/2025/000508/TS.

\appendix
\begin{widetext}

\setlength{\arrayrulewidth}{0.1mm}
\setlength{\tabcolsep}{8pt}
\renewcommand{\arraystretch}{1.4}
\begin{table*}[t]
\centering
\caption{Posterior constraints (68\% C.L.) on the Barrow entropy model parameters for the three IR cut-off prescriptions.
A dash indicates that the parameter is absent from the model; entries reported as inequalities denote 68\% C.L.\ upper or lower bounds on unconstrained (prior-dominated) parameters.
$\Delta\mathrm{AIC}$ and $\Delta\mathrm{BIC}$ are computed relative to the overall best-fitting model, Barrow\,+\,cut-off 3.}
\label{tab:barrow_results}
\begin{adjustbox}{max width=\textwidth}
\begin{tabular}{|l|c|c|c|c|c|c|c|c|c|c|c|c|}
\hline
Cut-off & $H_0$ & $\Omega_{m0}$ & $C$ & $\alpha$ & $\beta$ & $\gamma$ & $\Delta$ & $\Omega'_\mathrm{DE}(0)$ & $M_b$ & $\Omega_b$ & $\Delta\mathrm{AIC}$ & $\Delta\mathrm{BIC}$ \\
\hline
GO
& $67.63^{+0.67}_{-0.63}$
& $0.311^{+0.006}_{-0.006}$
& $0.543^{+0.057}_{-0.048}$
& $2.85^{+0.58}_{-0.51}$
& $1.43^{+0.30}_{-0.26}$
& ---
& $< 0.00432$
& ---
& $-19.443 \pm 0.016$
& $0.0486 \pm 0.0010$
& $148.07$ & $137.31$ \\
\hline
Cut-off 2
& $67.74^{+0.59}_{-0.62}$
& $0.310^{+0.006}_{-0.005}$
& $0.891^{+0.090}_{-0.078}$
& $1.06^{+0.21}_{-0.19}$
& $0.530^{+0.110}_{-0.097}$
& $> 6.88$
& $< 0.00433$
& ---
& $-19.442 \pm 0.016$
& $0.0484 \pm 0.0010$
& $150.59$ & $145.21$ \\
\hline
Cut-off 3
& $67.67^{+0.59}_{-0.62}$
& $0.310^{+0.006}_{-0.006}$
& $13.4^{+4.2}_{-6.1}$
& $0.50^{+0.26}_{-0.20}$
& $4.03^{+0.69}_{-1.13}$
& $1.21^{+0.21}_{-0.35}$
& $0.28^{+0.28}_{-0.20}$
& $-0.594^{+0.017}_{-0.016}$
& $-19.421^{+0.014}_{-0.015}$
& $0.0486^{+0.0009}_{-0.0010}$
& $0.00$ & $0.00$ \\
\hline
\end{tabular}
\end{adjustbox}
\end{table*}
\begin{table*}[t]
\centering
\caption{Same as Table~\ref{tab:barrow_results} for the $S_3$ (three-parameter) entropy.
$\Delta\mathrm{AIC}$ and $\Delta\mathrm{BIC}$ are computed relative to Barrow\,+\,cut-off 3, as in Table~\ref{tab:barrow_results}. ``Unc.'' denotes a prior-dominated parameter.}
\label{tab:S3_results}
\begin{adjustbox}{max width=\textwidth}
\begin{tabular}{|l|c|c|c|c|c|c|c|c|c|c|c|c|}
\hline
Cut-off & $H_0$ & $H_0''$ & $\Omega_{m0}$ & $C_r$ & $\alpha_r$ & $\gamma_r$ & $\alpha_{sr}$ & $\beta_s$ &  $M_b$ & $\Omega_b$ &$\Delta\mathrm{AIC}$ & $\Delta\mathrm{BIC}$ \\
\hline
GO
& $67.64^{+0.56}_{-0.60}$
& ---
& $0.311^{+0.0053}_{-0.0052}$
& $0.184^{+0.025}_{-0.025}$
& $1.797^{+0.190}_{-0.110}$
& ---
& $12.987^{+2.130}_{-2.086}$
& $1.212^{+0.113}_{-0.075}$
&$-19.421^{+0.014}_{-0.017}$
&$0.0486^{+0.0009}_{-0.0009}$
& $3.09$ & $24.61$ \\
\hline
Cut-off 2
& $67.97^{+0.60}_{-0.61}$
& ---
& $0.284^{+0.006}_{-0.005}$
& $0.273^{+0.046}_{-0.038}$
& $1.640^{+0.002}_{-0.001}$
& $2.626^{+81.647}_{-78.472}$
& $3.268^{+0.475}_{-0.434}$
& $2.283^{+0.658}_{-0.577}$
&$-19.443^{+0.015}_{-0.016}$
&$0.0502^{+0.0009}_{-0.0009}$
& $142.98$ & $169.88$ \\
\hline
Cut-off 3
& $68.02^{+0.67}_{-0.56}$
& \text{Unc.}
& $0.284 \pm 0.005$
& $0.172^{+0.013}_{-0.012}$
& $1.651^{+0.017}_{-0.008}$
& $-0.002^{+0.002}_{-0.003}$
& $6.514^{+0.850}_{-0.907}$
& $2.821^{+0.408}_{-0.354}$
&$-19.441^{+0.016}_{-0.014}$
&$0.0501^{+0.0009}_{-0.0010}$
& $144.73$ & $177.01$ \\
\hline
\end{tabular}
\end{adjustbox}
\end{table*}
\begin{table*}[t]
\centering
\caption{Same as Table~\ref{tab:barrow_results} for the $S_5$ (five-parameter) entropy.
$\Delta\mathrm{AIC}$ and $\Delta\mathrm{BIC}$ are computed relative to Barrow\,+\,cut-off 3, as in Table~\ref{tab:barrow_results}.}
\label{tab:S5_results}
\begin{adjustbox}{max width=\textwidth}
\begin{tabular}{|l|c|c|c|c|c|c|c|c|c|c|c|c|c|c|}
\hline
Cut-off & $H_0$ & $H_0''$ & $\Omega_{m0}$ & $C_r$ & $\alpha_r$ & $\gamma_r$ & $\alpha_{sr1}$ & $\alpha_{sr2}$ & $\beta_s$ & $\varepsilon_s$ & $M_b$ & $\Omega_b$ & $\Delta\mathrm{AIC}$ & $\Delta\mathrm{BIC}$ \\
\hline
GO
& $68.00^{+0.68}_{-0.73}$
& ---
& $0.284^{+0.007}_{-0.006}$
& $0.227^{+0.044}_{-0.041}$
& $1.640^{+0.002}_{-0.001}$
& ---
& $17.848^{+16.160}_{-12.431}$
& $10.830^{+11.743}_{-8.332}$
& $8.587^{+4.785}_{-2.908}$
& $25.814^{+20.412}_{-14.094}$
&$-19.441^{+0.018}_{-0.018}$
&$0.0502^{+0.0011}_{-0.0010}$
& $145.02$ & $177.31$ \\
\hline
Cut-off 2
& $68.00^{+0.74}_{-0.76}$
& ---
& $0.284 \pm 0.006$
& $0.518^{+0.124}_{-0.099}$
& $1.640^{+0.002}_{-0.001}$
& $10.099^{+46.796}_{-35.573}$
& $5.211^{+5.361}_{-3.706}$
& $4.950^{+5.197}_{-3.329}$
& $4.719^{+2.236}_{-1.620}$
& $27.289^{+18.983}_{-14.683}$
&$-19.442^{+0.018}_{-0.018}$
&$0.0502^{+0.0011}_{-0.0012}$
& $146.63$ & $184.30$ \\
\hline
Cut-off 3
& $67.46^{+0.71}_{-0.79}$
& $24.42^{+2.86}_{-2.32}$
& $0.312^{+0.007}_{-0.006}$
& $1.679^{+0.313}_{-0.231}$
& $0.334^{+0.035}_{-0.043}$
& $0.253^{+0.009}_{-0.008}$
& $3.545^{+1.735}_{-1.866}$
& $5.768^{+2.281}_{-1.442}$
& $11.886^{+1.181}_{-1.624}$
& $31.775^{+10.526}_{-14.617}$
&$-19.419^{+0.017}_{-0.021}$
&$0.0489^{+0.0012}_{-0.0011}$
& $17.61$ & $60.66$ \\
\hline
\end{tabular}
\end{adjustbox}
\end{table*}

\section{Closure relations for the $S_3$ and $S_5$ models}\label{app:H''}

In this appendix we give the explicit higher-derivative equations used to integrate the
$S_3$ and $S_5$ models. For all six combinations, the fractional dark energy density of
Secs.~\ref{sec:S3} and~\ref{sec:S5} can be written in the unified form
\begin{equation}\label{eq:unified}
    \Omega_\mathrm{DE} = \frac{C_r^2\,Y^2\,\mathcal{G}(Y)}{H^2}\,,
\end{equation}
where $Y$ encodes the cut-off,
\begin{align}
    Y_{\rm GO} &= \alpha_r H^2-(1+z)HH'\,,\qquad
    Y_{(2)} = Y_{\rm GO}+\frac{\gamma_r}{H^2}\,,\nonumber\\
    Y_{(3)} &= Y_{\rm GO}+\gamma_r(1+z)\left[HH'+(1+z)H'^2+(1+z)HH''\right],\label{eq:Ydef}
\end{align}
and $\mathcal{G}$ encodes the entropy,
\begin{align}
    \mathcal{G}_{S_3}(Y) &= \left(1+\frac{\alpha_{sr}}{Y}\right)^{\beta_s}-1\,,\nonumber\\
    \mathcal{G}_{S_5}(Y) &= \left[1+\frac{1}{\varepsilon_s}\arctan\frac{\alpha_{sr1}}{Y}\right]^{\beta_s}
    -\left[1+\frac{1}{\varepsilon_s}\arctan\frac{\alpha_{sr2}}{Y}\right]^{-\beta_s}.
\end{align}
For cut-off 3, $Y_{(3)}=A_{1}/H$ ($S_3$) and $Y_{(3)}=A_{2}/H$ ($S_5$), with $A_{1,2}$ defined in Eqs.~(\ref{S3ODEIIr}) and~(\ref{S5ODEIIr}).
Inserting Eq.~\eqref{eq:unified} into the Friedmann equation~\eqref{eq:Hubble} gives the
algebraic constraint
\begin{equation}
    C_r^2\,Y^2\,\mathcal{G}(Y) = H^2 - H_0^2(1+z)^3\,\Sigma\,,
\end{equation}
with $\Sigma$ defined in Eq.~\eqref{eq:Sigma_E}. Its derivative with respect to $z$ reads
$\mathcal{W}(Y)\,dY/dz = 2HH' - H_0^2(1+z)^2\left[3\Omega_{m0}+4(1+z)\Omega_{r0}\right]$, where
\begin{equation}
    \mathcal{W}(Y) \equiv C_r^2\left[2Y\mathcal{G}(Y)+Y^2\mathcal{G}'(Y)\right],
\end{equation}
and
\begin{align}
    \mathcal{G}_{S_3}'(Y) &= -\frac{\alpha_{sr}\beta_s}{Y^2}\left(1+\frac{\alpha_{sr}}{Y}\right)^{\beta_s-1},\nonumber\\
    \mathcal{G}_{S_5}'(Y) &= -\frac{\beta_s}{\varepsilon_s}\left\{\frac{\alpha_{sr1}}{Y^2+\alpha_{sr1}^2}\left[1+\frac{1}{\varepsilon_s}\arctan\frac{\alpha_{sr1}}{Y}\right]^{\beta_s-1}
    +\frac{\alpha_{sr2}}{Y^2+\alpha_{sr2}^2}\left[1+\frac{1}{\varepsilon_s}\arctan\frac{\alpha_{sr2}}{Y}\right]^{-\beta_s-1}\right\}.
\end{align}
Defining
\begin{equation}
    \mathcal{Q}(z) \equiv \frac{2HH' - H_0^2(1+z)^2\left[3\Omega_{m0}+4(1+z)\Omega_{r0}\right]}{\mathcal{W}(Y)}\,,
\end{equation}
and solving $dY/dz$ for the highest derivative, we obtain for the GO cut-off and cut-off 2
\begin{align}
    H''_{\rm GO} &= \frac{(2\alpha_r-1)HH'-(1+z)H'^2-\mathcal{Q}}{(1+z)H}\,,\label{eq:HppGO}\\
    H''_{(2)} &= \frac{(2\alpha_r-1)HH'-(1+z)H'^2-2\gamma_r H'/H^3-\mathcal{Q}}{(1+z)H}\,,\label{eq:Hpp2}
\end{align}
and for cut-off 3
\begin{equation}\label{eq:Hppp3}
    H'''_{(3)} = \frac{\mathcal{Q}-(2\alpha_r-1)HH'+(1+z)\left(H'^2+HH''\right)
    -\gamma_r\left[HH'+3(1+z)\left(H'^2+HH''\right)+3(1+z)^2H'H''\right]}{\gamma_r(1+z)^2H}\,,
\end{equation}
where in each case $\mathcal{Q}$ is evaluated with the corresponding $Y$ and $\mathcal{G}$.
Equations~\eqref{eq:HppGO}--\eqref{eq:Hppp3}, together with the boundary conditions discussed in
Secs.~\ref{sec:S3} and~\ref{sec:S5}, close the system for all six $S_3$ and $S_5$ models.

\section{Posterior distributions}\label{app:corner}
 
In this appendix we collect the corner plots showing the marginalized one- and two-dimensional posterior distributions for all entropy--cut-off combinations analyzed in this work.  In every panel the inner (outer) contours enclose the 68\% (95\%) credible regions, and dashed lines mark the median and the 16th/84th percentiles.  Parameters whose posterior is prior-dominated are flagged as upper or lower bounds (68\% C.L.) in the corresponding table.  The nuisance parameters $M_b$ and $\omega_b$ are included in the sampling but, where omitted from a corner plot, their posteriors are fully consistent across models.

\subsection{Barrow entropy}
 
\begin{figure}[H]
\centering
\includegraphics[width=0.5\textwidth]{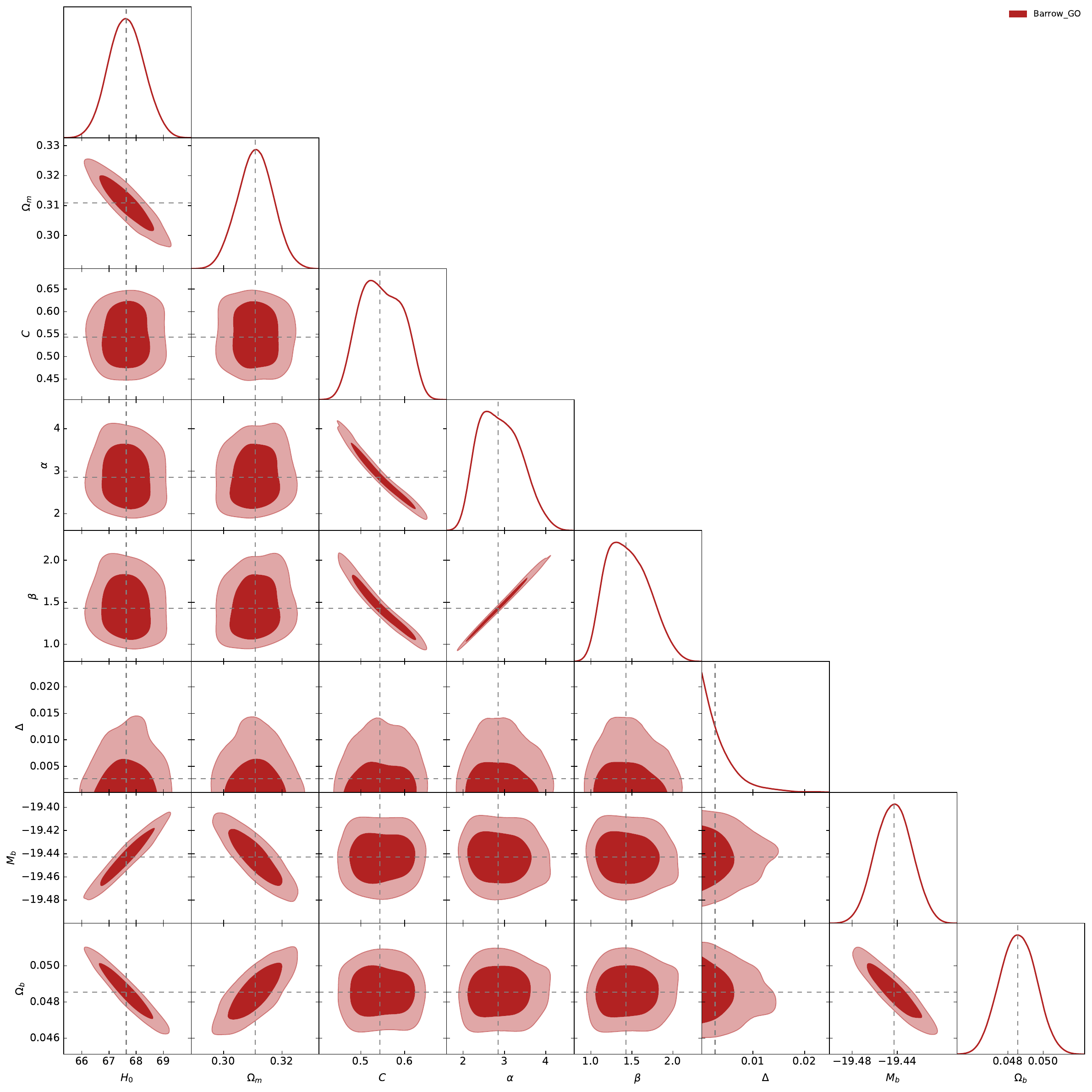}
\caption{Barrow\,+\,GO.
The Barrow exponent is tightly constrained to $\Delta < 0.00432$ (68\% C.L.), consistent with the Bekenstein--Hawking limit $\Delta = 0$.
The GO parameters $\alpha$ and $\beta$ are well determined and exhibit a pronounced positive degeneracy, which propagates into the holographic parameter $C$.
The base cosmological parameters $H_0$ and $\Omega_{m,0}$ are unaffected by the DE sector.}
\label{fig:corner_Barrow_GO}
\end{figure}
 
\begin{figure}[H]
\centering
\includegraphics[width=0.5\textwidth]{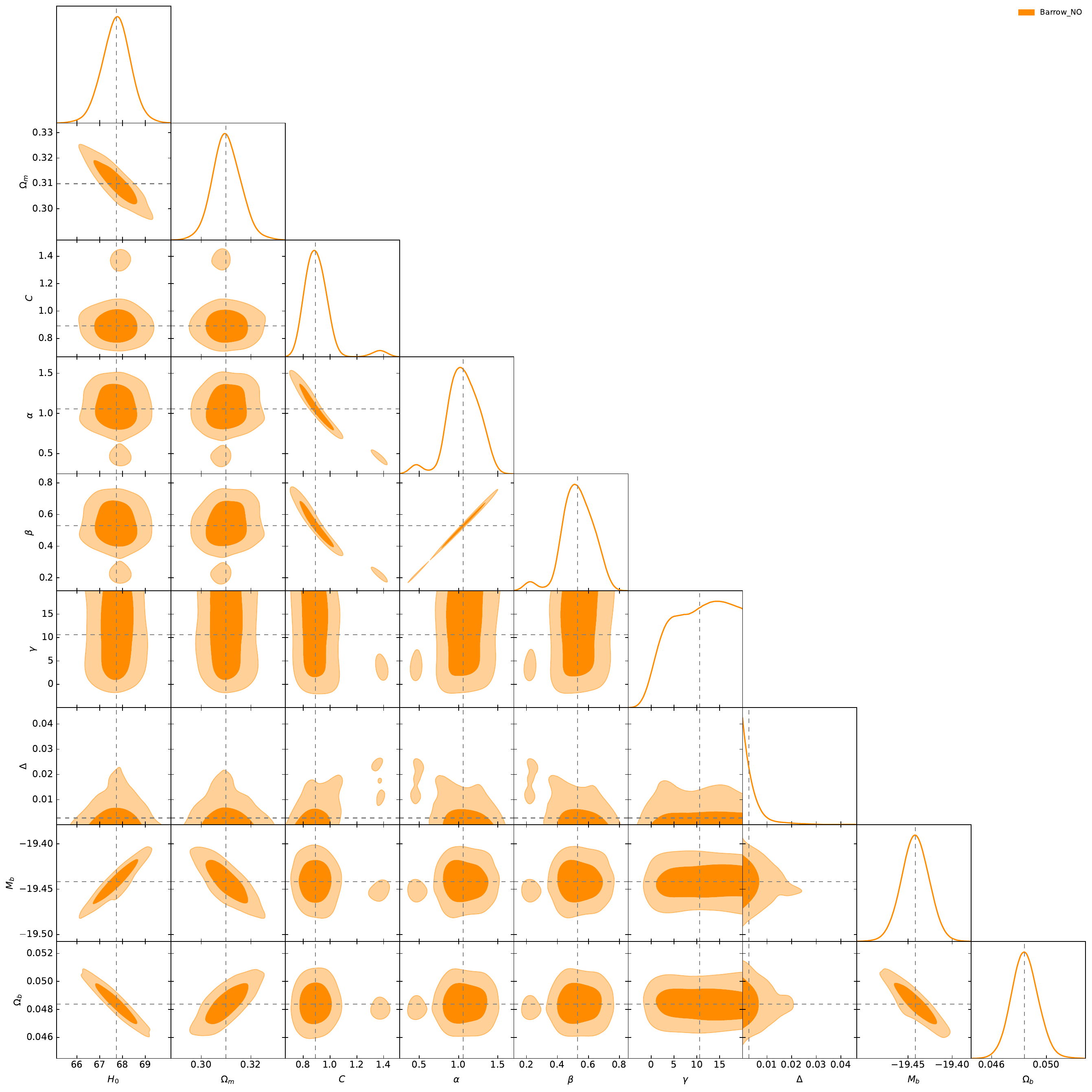}
\caption{Barrow\,+\,cut-off 2.
The additional parameter $\gamma$ is entirely unconstrained ($\gamma > 6.88$ at 68\% C.L.), with a posterior that mirrors the prior.
This confirms that the $\gamma\,H^{-2}$ correction becomes relevant only in the far future ($H \to 0$) and lies outside the sensitivity of current data.
All other parameters, including $\Delta < 0.00433$, remain consistent with the GO result, supporting the interpretation of GO as the data-preferred truncation.}
\label{fig:corner_Barrow_EGOI}
\end{figure}
 
\begin{figure}[H]
\centering
\includegraphics[width=0.5\textwidth]{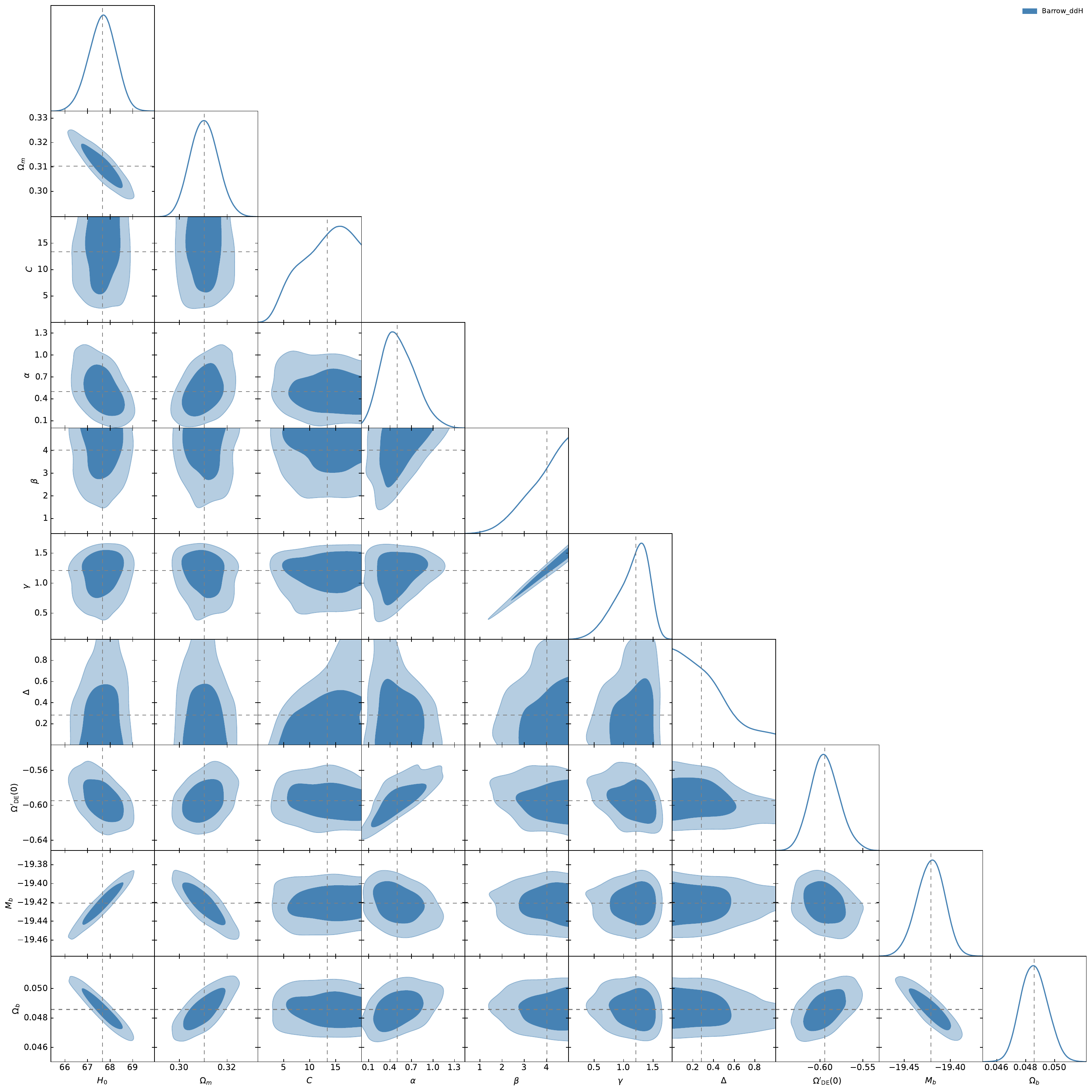}
\caption{Barrow\,+\,cut-off 3.
This is the only Barrow model in which $\Delta$ moves away from zero ($\Delta = 0.28^{+0.28}_{-0.20}$, compatible with zero at about $1.4\sigma$), enabled by the richer cut-off structure.
The shooting parameter $\Omega'_\mathrm{DE}(0)$ is tightly constrained to $-0.594^{+0.017}_{-0.016}$, indicating genuine sensitivity of the data to the second-order dynamics.
The holographic parameter $C$ is pushed to large values ($C = 13.4^{+4.2}_{-6.1}$), reflecting a strong $C$--$\Delta$ degeneracy.
This is the best-fitting model of the whole analysis.}
\label{fig:corner_Barrow_EGOII}
\end{figure}

\subsection{$S_3$ entropy}
\begin{figure}[H]
\centering
\includegraphics[width=0.4\textwidth]{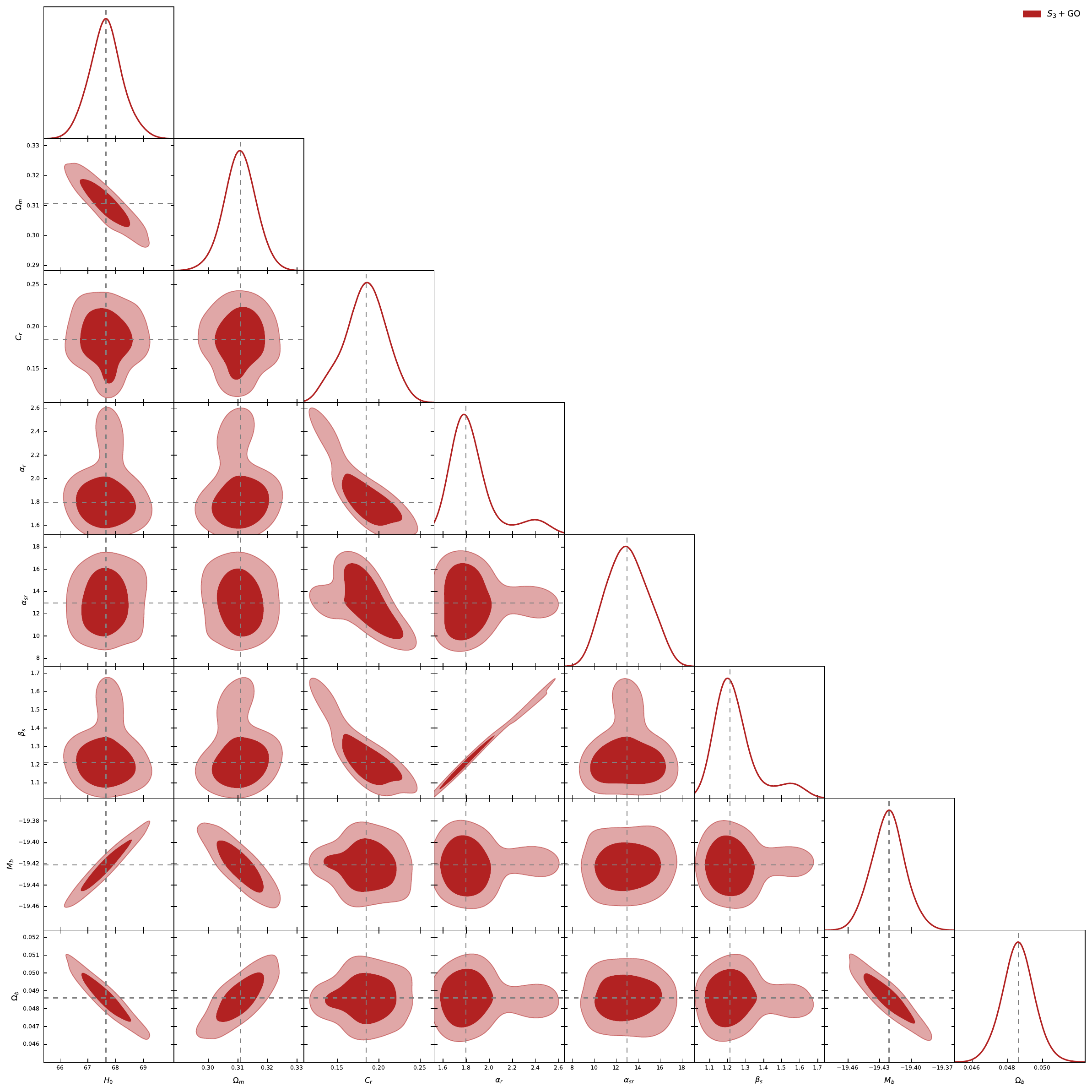}
\caption{$S_3$\,+\,GO.
The entropy parameters are well constrained, with $\alpha_{sr} = 12.987^{+2.130}_{-2.086}$ and $\beta_s = 1.212^{+0.113}_{-0.075}$.
The GO cut-off parameter $\alpha_r = 1.797^{+0.190}_{-0.110}$ shows a mild tail toward lower values.
This model achieves the best fit in the $S_3$ sector and the second-best overall ($\Delta\mathrm{AIC} = 3.09$), with a moderate BIC penalty from its larger number of parameters.}
\label{fig:corner_S3_GO}
\end{figure}
 
\begin{figure}[H]
\centering
\includegraphics[width=0.5\textwidth]{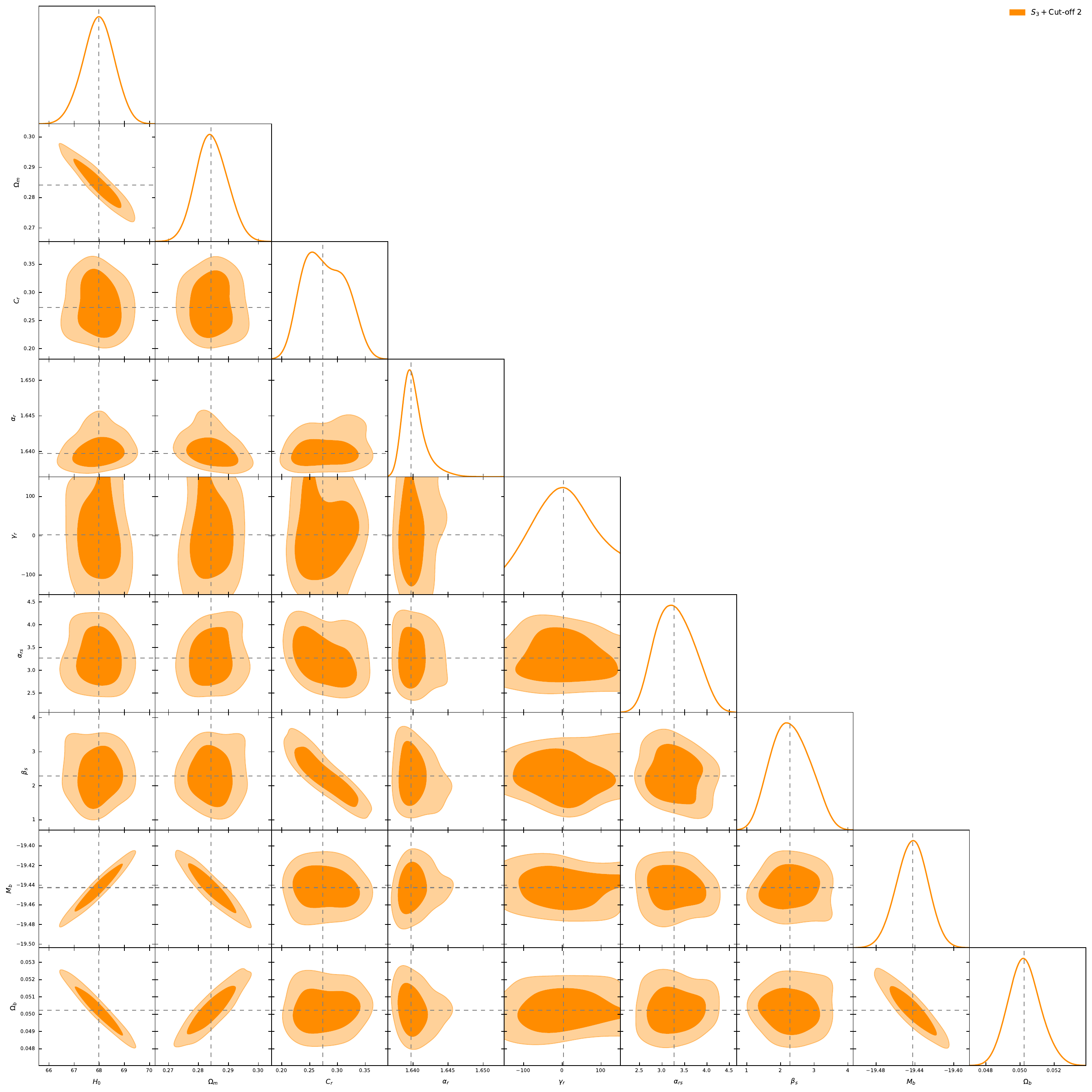}
\caption{$S_3$\,+\,cut-off 2.
The cut-off parameter $\gamma_r$ is almost unconstrained, analogous to the Barrow\,+\,cut-off 2 case.
The cut-off parameter $\alpha_r$ collapses to a narrow peak at $\alpha_r \simeq 1.640$ with sub-percent uncertainties, while the entropy parameters $\alpha_{sr}$ and $\beta_s$ shift substantially from their GO values, indicating a parameter-space reorganisation driven by the degeneracy with $\gamma_r$.
The large $\Delta\mathrm{AIC} = 142.98$ confirms that the additional parameter provides no improvement in fit.}
\label{fig:corner_S3_EGOI}
\end{figure}
 
\begin{figure}[H]
\centering
\includegraphics[width=0.5\textwidth]{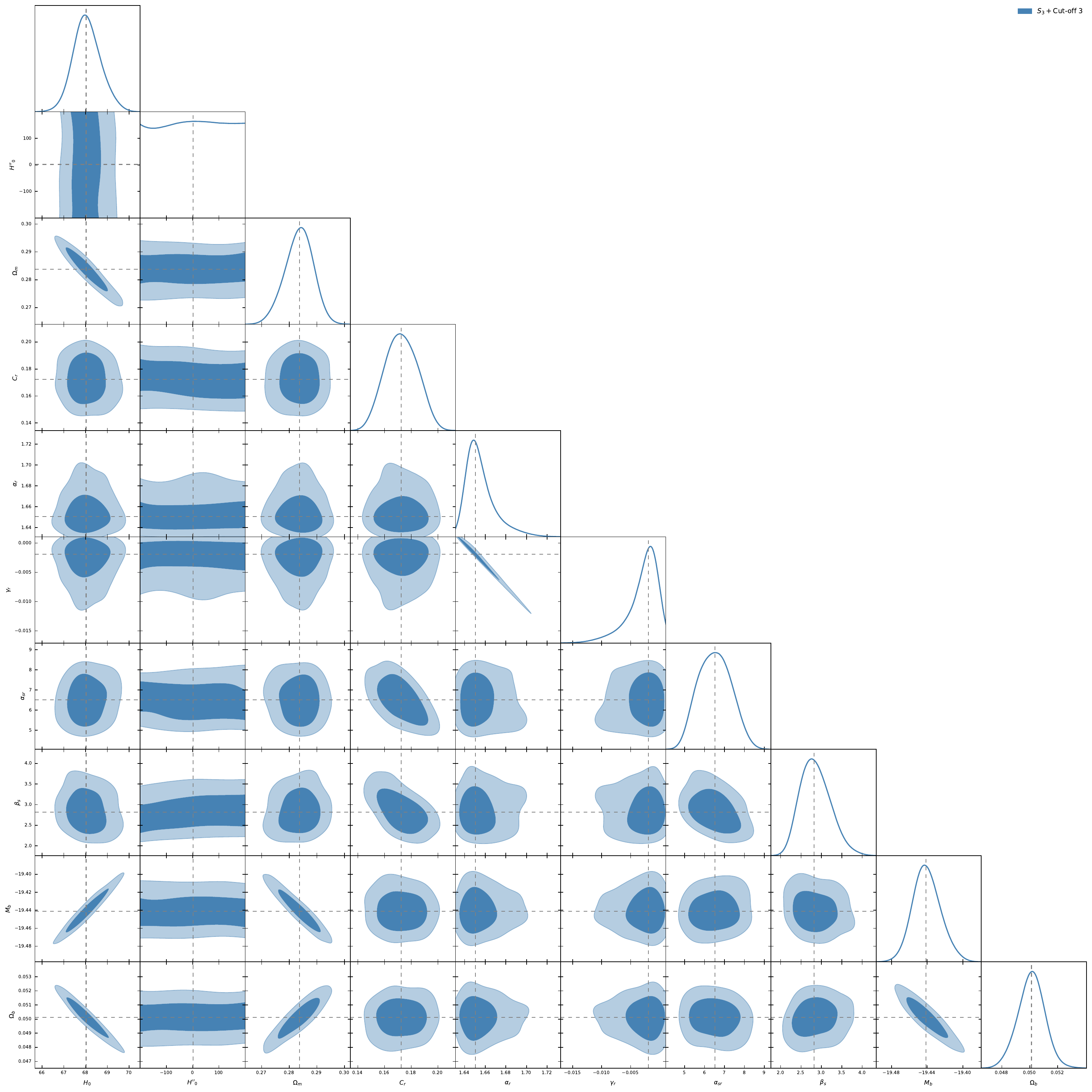}
\caption{$S_3$\,+\,cut-off 3.
The shooting parameter $H_0''$ is prior-dominated, with a broad posterior across the prior range, indicating that the data cannot determine the second initial condition.
The cut-off parameter $\gamma_r$ is constrained close to zero ($\gamma_r = -0.002^{+0.002}_{-0.003}$), consistent with the GO limit.
The entropy parameters $\alpha_{sr}$ and $\beta_s$ are well determined and take values intermediate between the GO and cut-off 2 results.}
\label{fig:corner_S3_EGOII}
\end{figure}

\subsection{$S_5$ entropy}
 
\begin{figure}[H]
\centering
\includegraphics[width=0.5\textwidth]{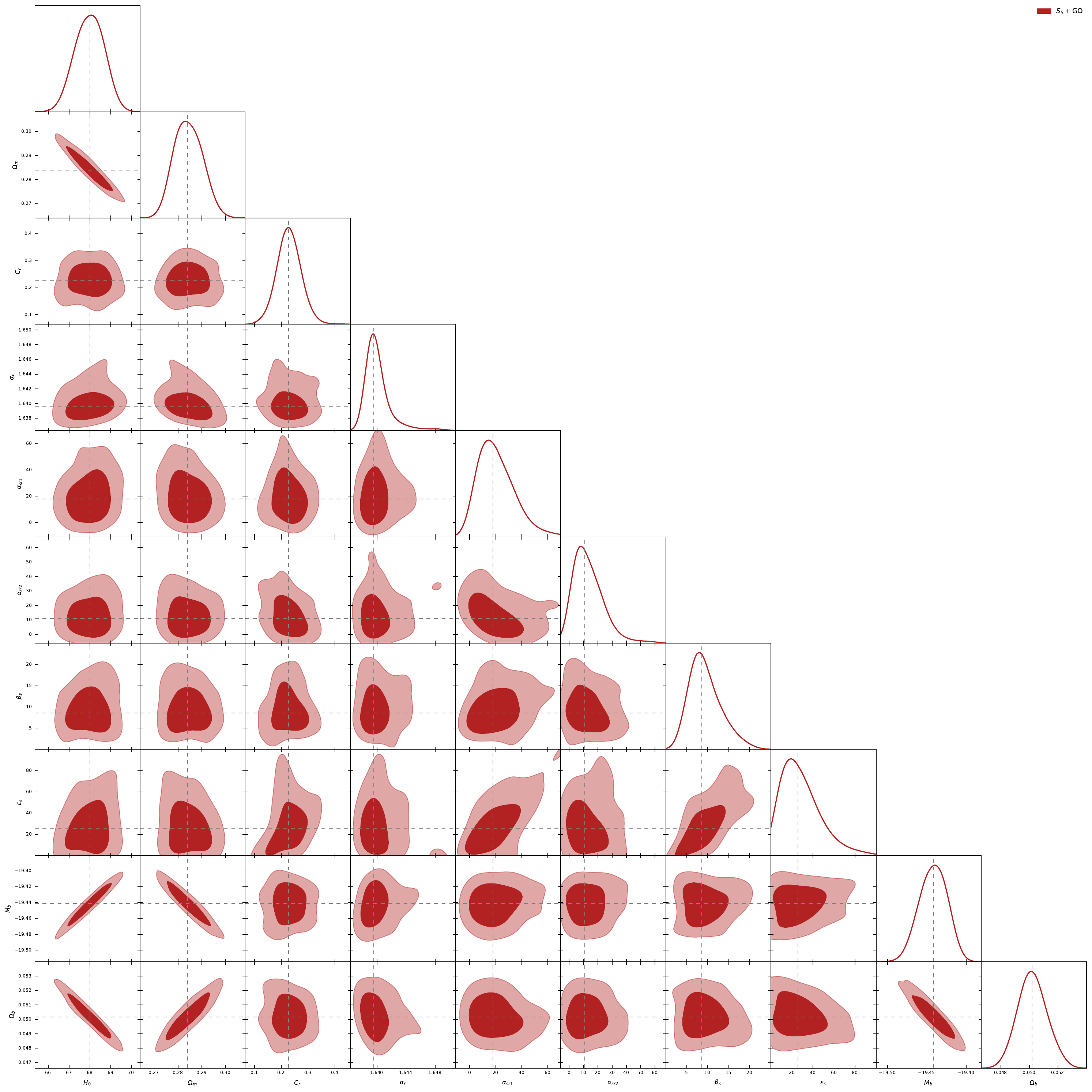}
\caption{$S_5$\,+\,GO.
The cut-off parameter is sharply constrained at $\alpha_r \simeq 1.640$, while the four entropy parameters $\{\alpha_{sr1}, \alpha_{sr2}, \beta_s, \varepsilon_s\}$ display broad, partially degenerate posteriors with large uncertainties.
In particular, $\alpha_{sr1}$ and $\alpha_{sr2}$ show significant mutual correlation and are only weakly constrained by the data, suggesting that the five-parameter entropy functional is over-parameterized for the current dataset.}
\label{fig:corner_S5_GO}
\end{figure}
 
\begin{figure}[H]
\centering
\includegraphics[width=0.5\textwidth]{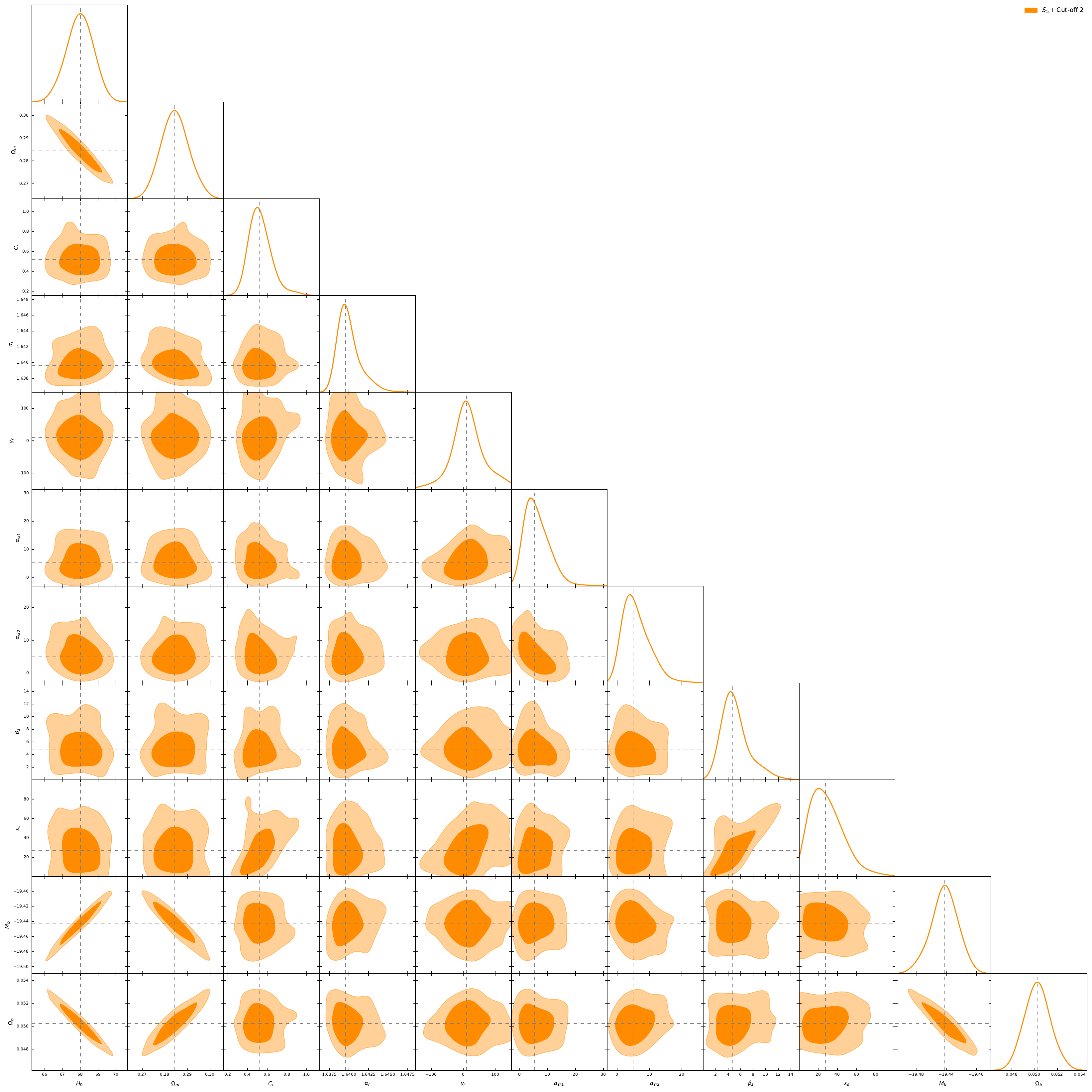}
\caption{$S_5$\,+\,cut-off 2.
The cut-off parameter $\gamma_r$ is again fully unconstrained ($\gamma_r = -10.099^{+46.796}_{-35.573}$), consistent with the pattern observed across all entropies.
The cut-off parameter $\alpha_r$ remains locked at $\alpha_r \simeq 1.640$, while $C_r$ shifts to a larger value ($C_r = 0.518$) compared to the GO case ($C_r = 0.227$).
The entropy parameters remain broadly distributed, confirming the over-parameterisation already seen in $S_5$\,+\,GO.}
\label{fig:corner_S5_EGOI}
\end{figure}
 
\begin{figure}[H]
\centering
\includegraphics[width=0.5\textwidth]{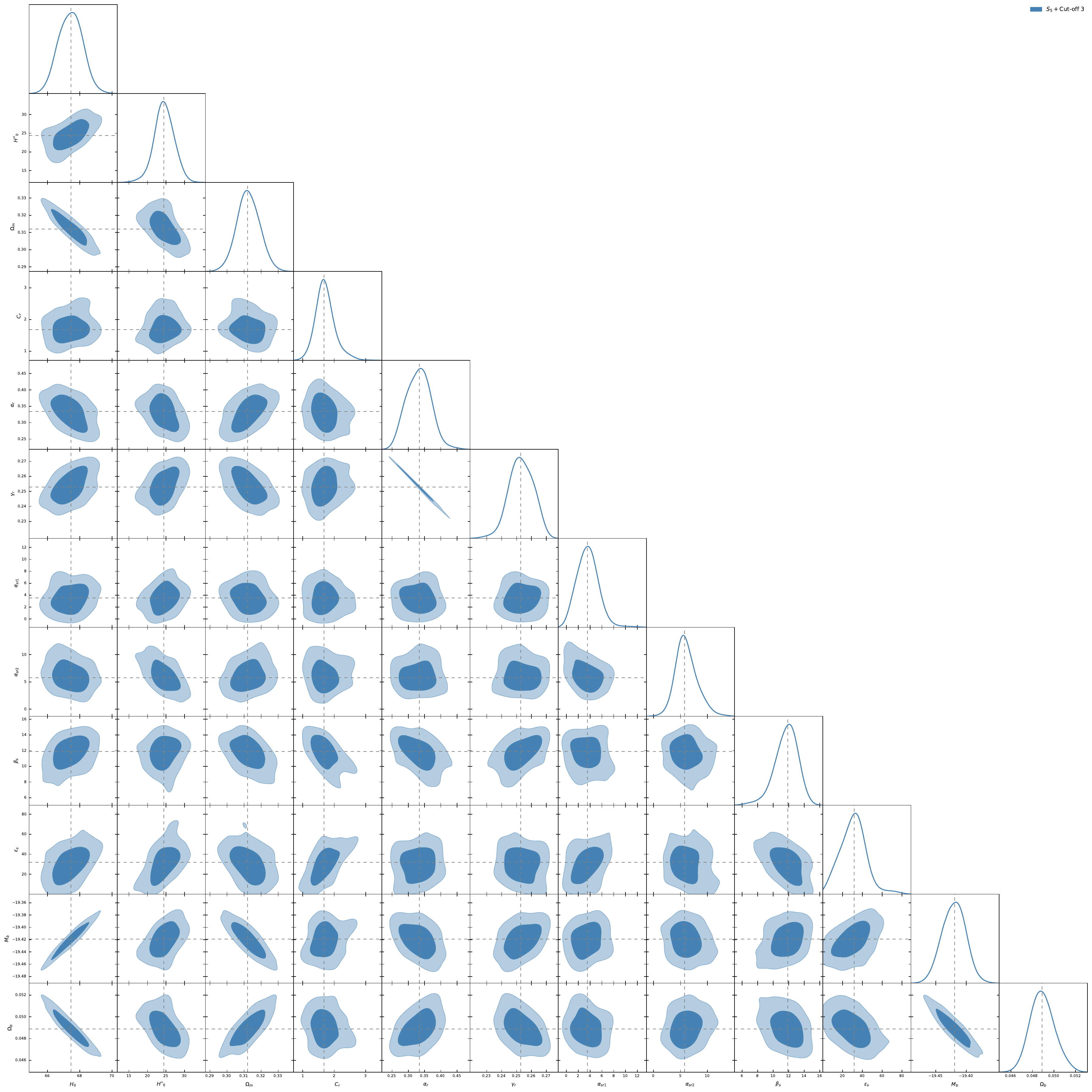}
\caption{$S_5$\,+\,cut-off 3.
This model is qualitatively distinct from all other $S_5$ cases: the shooting parameter $H_0''$ is well determined ($24.42^{+2.86}_{-2.32}$) and both cut-off parameters $\alpha_r$ and $\gamma_r$ are tightly constrained, with $\gamma_r = 0.253^{+0.009}_{-0.008}$ far from zero.
This model achieves the best fit in the $S_5$ sector ($\Delta\mathrm{AIC} = 17.61$) but remains penalized by its large parameter count ($\Delta\mathrm{BIC} = 60.66$).
The tight $\alpha_r$--$\gamma_r$ correlation visible in the corner plot reflects the interplay between the $H^2$ and $\ddot{H}/H$ contributions to the cut-off.}
\label{fig:corner_S5_EGOII}
\end{figure}

\section{The evolution of dark energy}\label{app:dark}
In this appendix, we present plots showing the evolution of the fractional dark energy density $\OmegaDE(z)$ and the EoS of dark energy $\omega_\mathrm{DE}(z)$ for $S_3$ entropy and $S_5$ entropy.

\begin{figure}[H]
\centering
\includegraphics[width=0.8\textwidth]{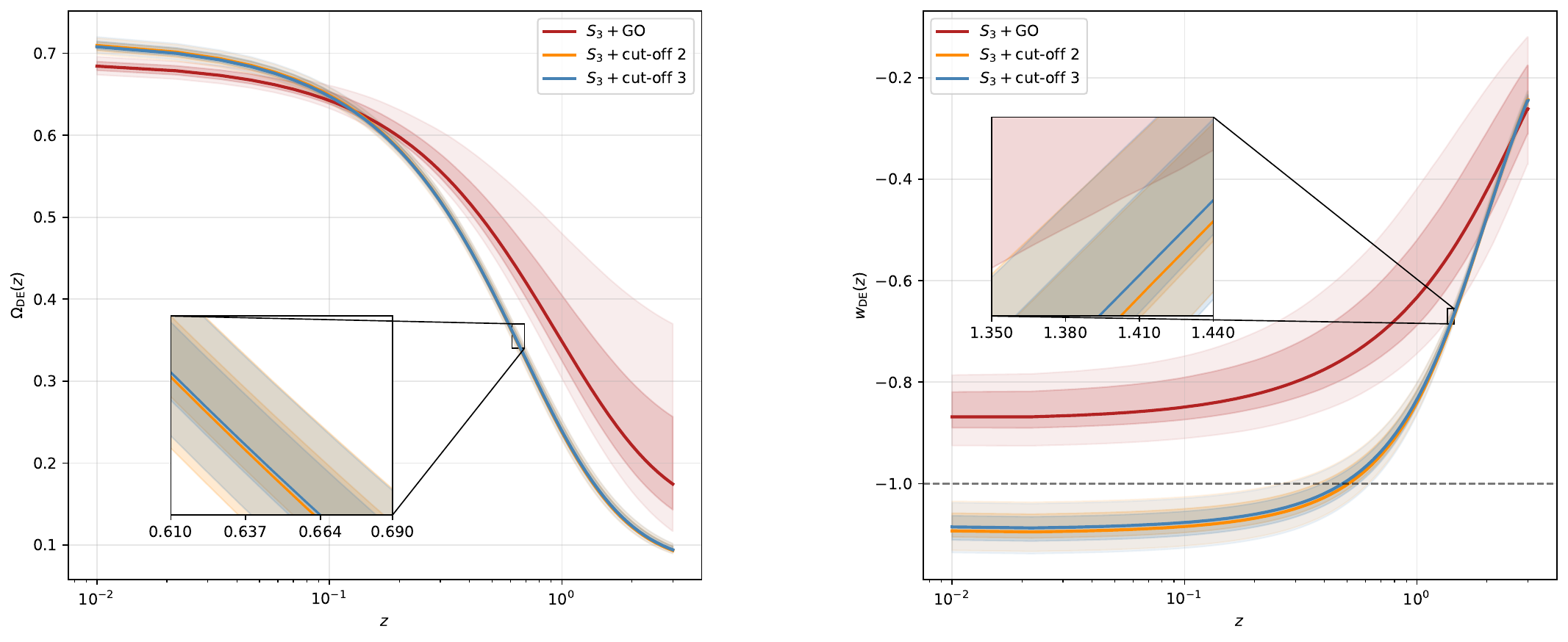}
\caption{The evolution of $\OmegaDE(z)$ and $\omega_\mathrm{DE}$ for $S_3$ entropy. The shaded regions indicate the 68\% C.L. and 95\% C.L. The evolution of $\OmegaDE(z)$ and $\omega_\mathrm{DE}$ of $S_3$\,+\,cut-off 2 and $S_3$\,+\,cut-off 3 are very similar. $S_3$\,+\,GO exhibits quintessence-like behavior, while $S_3$\,+\,cut-off 2 and $S_3$\,+\,cut-off 3 demonstrate a quintessence-to-phantom transition.}
\label{fig:S3_dark}
\end{figure}

\begin{figure}[H]
\centering
\includegraphics[width=0.8\textwidth]{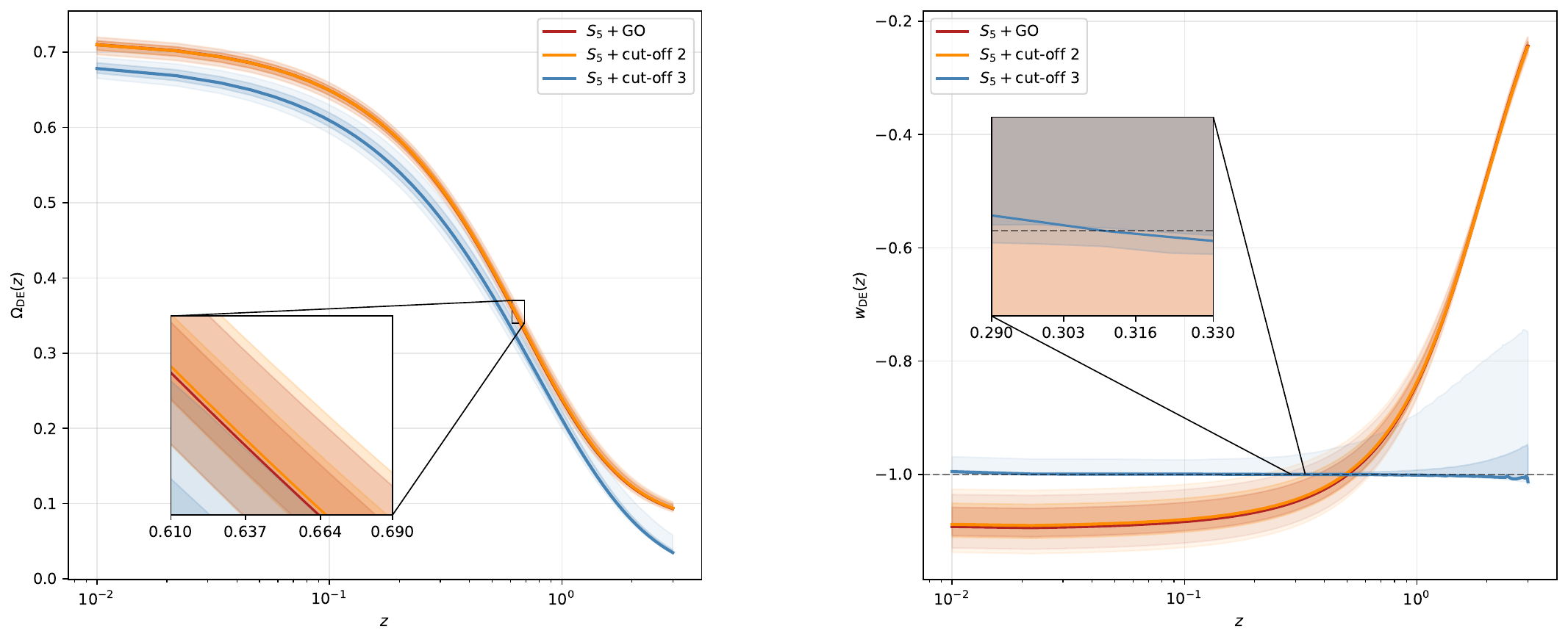}
\caption{The evolution of $\OmegaDE(z)$ and $\omega_\mathrm{DE}$ for $S_5$ entropy. The shaded regions indicate the 68\% C.L. and 95\% C.L. The evolution of $\OmegaDE(z)$ and $\omega_\mathrm{DE}$ of $S_5$\,+\,GO and $S_5$\,+\,cut-off 2 are very similar. $S_5$\,+\,GO and $S_5$\,+\,cut-off 2 demonstrate a quintessence-to-phantom transition, while $S_5$\,+\,cut-off 3 exhibits a mild phantom-to-quintessence transition.}
\label{fig:S5_dark}
\end{figure}

\end{widetext}
\phantomsection

\bibliographystyle{apsrev4-2}

\bibliography{Bibliography}

@article{Tyagi:2024cqp,
    author = "Tyagi, Udit K. and Haridasu, Sandeep and Basak, Soumen",
    title = "{Holographic and gravity-thermodynamic approaches in entropic
              cosmology: Bayesian assessment using late-time data}",
    eprint = "2406.07446",
    archivePrefix = "arXiv",
    primaryClass = "astro-ph.CO",
    doi = "10.1103/PhysRevD.110.063503",
    journal = "Phys. Rev. D",
    number = "6",
    pages = "063503",
    year = "2024"
}

@article{Tyagi:2025zov,
    author = "Tyagi, Udit K. and Haridasu, Sandeep and Basak, Soumen",
    title = "{Constraints on generalized gravity-thermodynamic cosmology from DESI DR2}",
    eprint = "2504.11308",
    archivePrefix = "arXiv",
    primaryClass = "astro-ph.CO",
    doi = "10.1103/7jkz-l38r",
    journal = "Phys. Rev. D",
    volume = "113",
    number = "6",
    pages = "063507",
    year = "2026"
}

@article{Odintsov:2024sbo,
    author = "Odintsov, Sergei D. and D'Onofrio, Simone and Paul, Tanmoy",
    title = "{Primordial gravitational waves in horizon cosmology and constraints on entropic parameters}",
    eprint = "2407.05855",
    archivePrefix = "arXiv",
    primaryClass = "gr-qc",
    doi = "10.1103/PhysRevD.110.043539",
    journal = "Phys. Rev. D",
    volume = "110",
    number = "4",
    pages = "043539",
    year = "2024"
}

@article{DOnofrio:2025bol,
    author = "D'Onofrio, Simone and Odintsov, Sergei and Paul, Tanmoy",
    title = "{Fitting NANOGrav 15-year data and ACT data with modified inflation in entropic cosmology}",
    eprint = "2510.20484",
    archivePrefix = "arXiv",
    primaryClass = "gr-qc",
    doi = "10.1103/ktt2-lm19",
    journal = "Phys. Rev. D",
    volume = "113",
    number = "4",
    pages = "043527",
    year = "2026"
}

@article{Odintsov:2025sew,
    author = "Odintsov, Sergei D. and D'Onofrio, Simone and Paul, Tanmoy",
    title = "{Generalized entropic dark energy with spatial curvature}",
    eprint = "2504.03470",
    archivePrefix = "arXiv",
    primaryClass = "gr-qc",
    doi = "10.1016/j.dark.2025.101920",
    journal = "Phys. Dark Univ.",
    volume = "48",
    pages = "101920",
    year = "2025"
}

@article{DOnofrio:2023jgc,
    author = "D'Onofrio, Simone",
    title = "{Holographic description of F(R) gravity coupled with axion dark matter}",
    eprint = "2306.16743",
    archivePrefix = "arXiv",
    primaryClass = "gr-qc",
    doi = "10.1142/S0219887823502249",
    journal = "Int. J. Geom. Meth. Mod. Phys.",
    volume = "20",
    number = "13",
    pages = "2350224",
    year = "2023"
}

@article{Oliveros:2022biu,
    author = "Oliveros, A. and Sabogal, M. A. and Acero, Mario A.",
    title = "{Barrow holographic dark energy with Granda{\textendash}Oliveros cutoff}",
    eprint = "2203.14464",
    archivePrefix = "arXiv",
    primaryClass = "gr-qc",
    doi = "10.1140/epjp/s13360-022-02994-z",
    journal = "Eur. Phys. J. Plus",
    volume = "137",
    number = "7",
    pages = "783",
    year = "2022"
}

@article{Cohen:1998zx,
    author = "Cohen, Andrew G. and Kaplan, David B. and Nelson, Ann E.",
    title = "{Effective field theory, black holes, and the cosmological constant}",
    eprint = "hep-th/9803132",
    archivePrefix = "arXiv",
    reportNumber = "BUHEP-98-7, DOE-ER-40561-358, INT-98-00-6, UW-PT-97-24",
    doi = "10.1103/PhysRevLett.82.4971",
    journal = "Phys. Rev. Lett.",
    volume = "82",
    pages = "4971--4974",
    year = "1999"
}

@article{Wang:2016och,
    author = "Wang, Shuang and Wang, Yi and Li, Miao",
    title = "{Holographic Dark Energy}",
    eprint = "1612.00345",
    archivePrefix = "arXiv",
    primaryClass = "astro-ph.CO",
    doi = "10.1016/j.physrep.2017.06.003",
    journal = "Phys. Rept.",
    volume = "696",
    pages = "1--57",
    year = "2017"
}

@article{Nojiri:2005pu,
    author = "Nojiri, Shin'ichi and Odintsov, Sergei D.",
    title = "{Unifying phantom inflation with late-time acceleration: Scalar
              phantom-non-phantom transition model and generalized holographic
              dark energy}",
    eprint = "hep-th/0506212",
    archivePrefix = "arXiv",
    doi = "10.1007/s10714-006-0301-6",
    journal = "Gen. Rel. Grav.",
    volume = "38",
    pages = "1285--1304",
    year = "2006"
}

@article{Nojiri:2017opc,
    author = "Nojiri, Shin'ichi and Odintsov, Sergei D.",
    title = "{Covariant generalized holographic dark energy and accelerating
              universe}",
    eprint = "1703.06372",
    archivePrefix = "arXiv",
    primaryClass = "hep-th",
    doi = "10.1140/epjc/s10052-017-5097-x",
    journal = "Eur. Phys. J. C",
    volume = "77",
    pages = "528",
    year = "2017"
}

@article{Granda:2008dk,
    author = "Granda, L. N. and Oliveros, A.",
    title = "{Infrared cut-off proposal for the holographic density}",
    eprint = "0810.3149",
    archivePrefix = "arXiv",
    primaryClass = "gr-qc",
    doi = "10.1016/j.physletb.2008.10.017",
    journal = "Phys. Lett. B",
    volume = "669",
    pages = "275--279",
    year = "2008"
}

@article{Granda:2009fh,
    author = "Granda, L. N. and Oliveros, A.",
    title = "{New infrared cut-off for the holographic scalar fields models of dark energy}",
    eprint = "0810.3663",
    archivePrefix = "arXiv",
    primaryClass = "gr-qc",
    doi = "10.1016/j.physletb.2008.12.025",
    journal = "Phys. Lett. B",
    volume = "671",
    pages = "199--202",
    year = "2009"
}

@article{Yarahmadi:2024mnras,
    author = "Yarahmadi, Muhammad and Salehi, Amin",
    title = "{Alleviating the Hubble tension using the Barrow holographic dark
              energy cosmology with {Granda--Oliveros} IR cut-off}",
    doi = "10.1093/mnras/stae2257",
    journal = "Mon. Not. Roy. Astron. Soc.",
    volume = "534",
    number = "4",
    pages = "3055--3067",
    year = "2024"
}

@article{Luciano:2025jcap,
    author = "Luciano, Giuseppe Gaetano and Paliathanasis, Andronikos
              and Saridakis, Emmanuel N.",
    title = "{Barrow and Tsallis entropies after the {DESI DR2 BAO} data}",
    eprint = "2504.12205",
    archivePrefix = "arXiv",
    primaryClass = "astro-ph.CO",
    doi = "10.1088/1475-7516/2025/09/013",
    journal = "JCAP",
    volume = "09",
    pages = "013",
    year = "2025"
}

@article{Luciano:2026jheap,
    author = "Luciano, Giuseppe Gaetano and Paliathanasis, Andronikos
              and Saridakis, Emmanuel N.",
    title = "{Constraints on Barrow and Tsallis holographic dark energy from
              {DESI DR2 BAO} data}",
    eprint = "2506.03019",
    archivePrefix = "arXiv",
    primaryClass = "astro-ph.CO",
    doi = "10.1016/j.jheap.2025.100427",
    journal = "JHEAp",
    volume = "49",
    pages = "100427",
    year = "2026"
}

@article{Chattopadhyay:2025epjc,
       author = {{Bekova}, Guldana and {Altaibayeva}, Aziza and {Ualikhanova}, Ulbossyn and {Chattopadhyay}, Surajit},
        title = "{Cosmology of barrow holographic dark energy with generalized infrared cutoffs and validity of the generalized second law of thermodynamics}",
      journal = {European Physical Journal C},
         year = 2025,
        month = oct,
       volume = {85},
       number = {10},
          eid = {1135},
        pages = {1135},
          doi = {10.1140/epjc/s10052-025-14843-1},
       adsurl = {https://ui.adsabs.harvard.edu/abs/2025EPJC...85.1135B}
}

@misc{Singh:2026ihe,
    author = "Singh, Swapnil Kumar",
    title = "{Oscillatory freeze from inertial holographic dark energy}",
    eprint = "2601.07007",
    archivePrefix = "arXiv",
    primaryClass = "gr-qc",
    year = "2026"
}

@article{Cai:2008wt,
    author = "Cai, Rong-Gen and Hu, Bin and Zhang, Yi",
    title = "{Holography, UV/IR relation, causal entropy bound
              and dark energy}",
    eprint = "0812.4504",
    archivePrefix = "arXiv",
    primaryClass = "hep-th",
    doi = "10.1088/0253-6102/51/5/34",
    journal = "Commun. Theor. Phys.",
    volume = "51",
    pages = "954--960",
    year = "2009"
}

@article{Mahmoudifard:2024epjc,
    author = "Mahmoudifard, Mojtaba and Salehi, Amin
              and Sepahvand, Reza",
    title = "{Constraining neutrino masses in the Barrow holographic
              dark energy model with {Granda--Oliveros} IR cutoff}",
    doi = "10.1140/epjc/s10052-024-13466-2",
    journal = "Eur. Phys. J. C",
    volume = "84",
    pages = "1099",
    year = "2024"
}

@article{Li:2004rb,
    author = "Li, Miao",
    title = "{A Model of holographic dark energy}",
    journal = "Phys. Lett. B",
    volume = "603",
    pages = "1--5",
    year = "2004",
    doi = "10.1016/j.physletb.2004.10.014",
    eprint = "hep-th/0403127"
}

@article{Hsu:2004ri,
    author = "Hsu, Stephen D. H.",
    title = "{Entropy bounds and dark energy}",
    journal = "Phys. Lett. B",
    volume = "594",
    pages = "13--16",
    year = "2004",
    doi = "10.1016/j.physletb.2004.05.020",
    eprint = "hep-th/0403052"
}

@article{Barrow:2020tzx,
    author = "Barrow, John D.",
    title = "{The Area of a Rough Black Hole}",
    journal = "Phys. Lett. B",
    volume = "808",
    pages = "135643",
    year = "2020",
    doi = "10.1016/j.physletb.2020.135643",
    eprint = "2004.09444"
}

@article{Saridakis:2020zol,
    author = "Saridakis, Emmanuel N.",
    title = "{Barrow holographic dark energy}",
    journal = "Phys. Rev. D",
    volume = "102",
    pages = "123525",
    year = "2020",
    doi = "10.1103/PhysRevD.102.123525",
    eprint = "2005.04115"
}

@article{Anagnostopoulos:2020ctz,
    author = "Anagnostopoulos, Fotios K. and Basilakos, Spyros and Saridakis, Emmanuel N.",
    title = "{Observational constraints on Barrow holographic dark energy}",
    journal = "Eur. Phys. J. C",
    volume = "80",
    pages = "826",
    year = "2020",
    doi = "10.1140/epjc/s10052-020-8360-5",
    eprint = "2005.10302"
}

@article{Leon:2021wyx,
    author = "Leon, Genly and others",
    title = "{Barrow entropy cosmology: an observational approach with a hint of stability analysis}",
    journal = "JCAP",
    volume = "12",
    pages = "032",
    year = "2021",
    doi = "10.1088/1475-7516/2021/12/032",
    eprint = "2108.10998"
}

@article{Nojiri:2022aof,
    author = "Nojiri, Shin'ichi and Odintsov, Sergei D. and Paul, Tanmoy",
    title = "{Early and late universe holographic cosmology from a new generalized entropy}",
    journal = "Phys. Lett. B",
    volume = "831",
    pages = "137189",
    year = "2022",
    doi = "10.1016/j.physletb.2022.137189",
    eprint = "2205.08876"
}

@article{Nojiri:2022dkr,
    author = "Nojiri, Shin'ichi and Odintsov, Sergei D. and Paul, Tanmoy",
    title = "{Modified cosmological perturbations and the three-parameter generalized entropy}",
    journal = "Phys. Rev. D",
    volume = "107",
    pages = "044033",
    year = "2023",
    doi = "10.1103/PhysRevD.107.044033",
    eprint = "2212.06767"
}

@article{Tsallis:2012js,
    author = "Tsallis, Constantino and Cirto, Leonardo J. L.",
    title = "{Black hole thermodynamical entropy}",
    journal = "Eur. Phys. J. C",
    volume = "73",
    pages = "2487",
    year = "2013",
    doi = "10.1140/epjc/s10052-013-2487-6"
}

@inproceedings{renyi1961measures,
  title={On measures of entropy and information},
  author={R{\'e}nyi, Alfr{\'e}d},
  booktitle={Proceedings of the fourth Berkeley symposium on mathematical statistics and probability, volume 1: contributions to the theory of statistics},
  volume={4},
  pages={547--562},
  year={1961},
  organization={University of California Press}
}

@article{Kaniadakis:2005zk,
    author = "Kaniadakis, Giorgio",
    title = "{Statistical mechanics in the context of special relativity. II}",
    journal = "Phys. Rev. E",
    volume = "72",
    pages = "036108",
    year = "2005",
    doi = "10.1103/PhysRevE.72.036108",
    eprint = "cond-mat/0507311"
}

@article{SayahianJahromi:2018irq,
    author = "Sayahian Jahromi, A. and Moosavi, S. A. and Moradpour, H. and Morais Graca, J. P. and Lobo, I. P. and Salako, I. G. and Jawad, A.",
    title = "{Generalized entropy formalism and a new holographic dark energy model}",
    journal = "Phys. Lett. B",
    volume = "780",
    pages = "21--24",
    year = "2018",
    doi = "10.1016/j.physletb.2018.02.052",
    eprint = "1802.07722"
}

@article{Odintsov:2022qnn,
    author = "Odintsov, Sergei D. and Paul, Tanmoy",
    title = "{A non-singular generalized entropy and its implications on bounce cosmology}",
    eprint = "2212.05531",
    archivePrefix = "arXiv",
    primaryClass = "gr-qc",
    doi = "10.1016/j.dark.2022.101159",
    journal = "Phys. Dark Univ.",
    volume = "39",
    pages = "101159",
    year = "2023"
}

@article{Brout_2022,
   title={The Pantheon+ Analysis: Cosmological Constraints},
   volume={938},
   ISSN={1538-4357},
   url={http://dx.doi.org/10.3847/1538-4357/ac8e04},
   DOI={10.3847/1538-4357/ac8e04},
   number={2},
   journal={The Astrophysical Journal},
   publisher={American Astronomical Society},
   author={Brout, Dillon and Scolnic, Dan and Popovic, Brodie and Riess, Adam G. and Carr, Anthony and Zuntz, Joe and Kessler, Rick and Davis, Tamara M. and Hinton et.al.},
   year={2022},
   month=oct, pages={110} }

@article{DESI:2025zgx,
    author = "Abdul Karim, M. and others",
    collaboration = "DESI",
    title = "{DESI DR2 results. II. Measurements of baryon acoustic oscillations and cosmological constraints}",
    eprint = "2503.14738",
    archivePrefix = "arXiv",
    primaryClass = "astro-ph.CO",
    reportNumber = "FERMILAB-PUB-25-0169-PPD",
    doi = "10.1103/tr6y-kpc6",
    journal = "Phys. Rev. D",
    volume = "112",
    number = "8",
    pages = "083515",
    year = "2025"
}

@article{Brieden_2023,
   title={A tale of two (or more) h’s},
   volume={2023},
   ISSN={1475-7516},
   url={http://dx.doi.org/10.1088/1475-7516/2023/04/023},
   DOI={10.1088/1475-7516/2023/04/023},
   number={04},
   journal={Journal of Cosmology and Astroparticle Physics},
   publisher={IOP Publishing},
   author={Brieden, Samuel and Gil-Marín, Héctor and Verde, Licia},
   year={2023},
   month=apr, pages={023} }

@article{Lemos:2023xhs,
    author = "Lemos, Pablo and Lewis, Antony",
    title = "{CMB constraints on the early Universe independent of late-time cosmology}",
    eprint = "2302.12911",
    archivePrefix = "arXiv",
    primaryClass = "astro-ph.CO",
    doi = "10.1103/PhysRevD.107.103505",
    journal = "Phys. Rev. D",
    volume = "107",
    number = "10",
    pages = "103505",
    year = "2023"
}

@article{Planck_2020,
   title={Planck2018 results: VI. Cosmological parameters},
   volume={641},
   ISSN={1432-0746},
   url={http://dx.doi.org/10.1051/0004-6361/201833910},
   DOI={10.1051/0004-6361/201833910},
   journal={Astronomy \& Astrophysics},
   publisher={EDP Sciences},
   author={Aghanim, N. and Akrami, Y. and Ashdown, M. and Aumont, J. and Baccigalupi, C. and Ballardini, M. and Banday, A. J. and Barreiro, R. B. andothers},
   year={2020},
   month=sep, pages={A6} }

@article{Hu_1996,
   title={Small‐Scale Cosmological Perturbations: An Analytic Approach},
   volume={471},
   ISSN={1538-4357},
   url={http://dx.doi.org/10.1086/177989},
   DOI={10.1086/177989},
   number={2},
   journal={The Astrophysical Journal},
   publisher={American Astronomical Society},
   author={Hu, Wayne and Sugiyama, Naoshi},
   year={1996},
   month=nov, pages={542–570} }

@article{Fixsen:2009ug,
    author = "Fixsen, D. J.",
    title = "{The Temperature of the Cosmic Microwave Background}",
    eprint = "0911.1955",
    archivePrefix = "arXiv",
    primaryClass = "astro-ph.CO",
    doi = "10.1088/0004-637X/707/2/916",
    journal = "Astrophys. J.",
    volume = "707",
    pages = "916--920",
    year = "2009"
}

@article{padilla2021cosmological,
  title={Cosmological parameter inference with Bayesian statistics},
volume={7},
   ISSN={2218-1997},
   url={http://dx.doi.org/10.3390/universe7070213},
   DOI={10.3390/universe7070213},
   number={7},
   journal={Universe},
   publisher={MDPI AG},
   author={Padilla, Luis E. and Tellez, Luis O. and Escamilla, Luis A. and Vazquez, Jose Alberto},
   year={2021},
   month=jun, pages={213} 
}

@article{Foreman-Mackey2019, 
doi = {10.21105/joss.01864}, 
url = {https://doi.org/10.21105/joss.01864}, 
year = {2019}, 
publisher = {The Open Journal}, 
volume = {4}, 
number = {43}, 
pages = {1864}, 
author = {Foreman-Mackey, Daniel and Farr, Will M. and Sinha, Manodeep and Archibald, Anne M. and Hogg, David W. and Sanders, Jeremy S. and Zuntz, Joe and Williams, Peter K. g. and Nelson, Andrew R. j. and de Val-Borro, Miguel and Erhardt, Tobias and Pashchenko, Ilya and Pla, Oriol Abril}, 
title = {emcee v3: A Python ensemble sampling toolkit for affine-invariant MCMC}, 
journal = {Journal of Open Source Software} 
}

@article{Riess:1998cb,
    author = "Riess, Adam G. and others",
    collaboration = "Supernova Search Team",
    title = "{Observational evidence from supernovae for an accelerating universe and a cosmological constant}",
    eprint = "astro-ph/9805201",
    archivePrefix = "arXiv",
    doi = "10.1086/300499",
    journal = "Astron. J.",
    volume = "116",
    pages = "1009--1038",
    year = "1998"
}

@article{Perlmutter:1999np,
    author = "Perlmutter, S. and others",
    collaboration = "Supernova Cosmology Project",
    title = "{Measurements of $\Omega$ and $\Lambda$ from 42 High Redshift Supernovae}",
    eprint = "astro-ph/9812133",
    archivePrefix = "arXiv",
    reportNumber = "LBNL-41801, LBL-41801",
    doi = "10.1086/307221",
    journal = "Astrophys. J.",
    volume = "517",
    pages = "565--586",
    year = "1999"
}

@article{Liddle:2007fy,
    author  = "Liddle, Andrew R.",
    title   = "{Information criteria for astrophysical model selection}",
    eprint  = "astro-ph/0701113",
    archivePrefix = "arXiv",
    doi     = "10.1111/j.1745-3933.2007.00306.x",
    journal = "Mon. Not. Roy. Astron. Soc.",
    volume  = "377",
    pages   = "L74--L78",
    year    = "2007"
}

@article{Akaike:1974,
    author  = "Akaike, Hirotugu",
    title   = "{A new look at the statistical model identification}",
    journal = "IEEE Trans. Automat. Contr.",
    volume  = "19",
    number  = "6",
    pages   = "716--723",
    year    = "1974",
    doi     = "10.1109/TAC.1974.1100705"
}

@article{Schwarz:1978,
    author  = "Schwarz, Gideon",
    title   = "{Estimating the dimension of a model}",
    journal = "Annals Statist.",
    volume  = "6",
    number  = "2",
    pages   = "461--464",
    year    = "1978",
    doi     = "10.1214/aos/1176344136"
}

@article{Nojiri:2021iko,
    author  = "Nojiri, Shin'ichi and Odintsov, Sergei D. and Paul, Tanmoy",
    title   = "{Different faces of generalized holographic dark energy}",
    eprint  = "2105.08438",
    archivePrefix = "arXiv",
    primaryClass  = "gr-qc",
    doi     = "10.3390/sym13060928",
    journal = "Symmetry",
    volume  = "13",
    number  = "6",
    pages   = "928",
    year    = "2021"
}

@article{Nojiri:2022cez,
    author  = "Nojiri, Shin'ichi and Odintsov, Sergei D. and Paul, Tanmoy",
    title   = "{Barrow entropic dark energy: A member of generalized
               holographic dark energy family}",
    eprint  = "2112.10159",
    archivePrefix = "arXiv",
    primaryClass  = "gr-qc",
    doi     = "10.1016/j.physletb.2021.136844",
    journal = "Phys. Lett. B",
    volume  = "825",
    pages   = "136844",
    year    = "2022"
}

@article{Gao:2007ep,
    author  = "Gao, Changjun and Wu, Fengquan and Chen, Xuelei and
               Shen, You-Gen",
    title   = "{A Holographic Dark Energy Model from Ricci Scalar
               Curvature}",
    eprint  = "0712.1394",
    archivePrefix = "arXiv",
    primaryClass  = "astro-ph",
    doi     = "10.1103/PhysRevD.79.043511",
    journal = "Phys. Rev. D",
    volume  = "79",
    pages   = "043511",
    year    = "2009"
}

@article{DAgostino:2019wko,
    author  = "D'Agostino, Rocco",
    title   = "{Holographic dark energy from nonadditive entropy:
               cosmological perturbations and observational constraints}",
    eprint  = "1903.03836",
    archivePrefix = "arXiv",
    primaryClass  = "gr-qc",
    doi     = "10.1103/PhysRevD.99.103524",
    journal = "Phys. Rev. D",
    volume  = "99",
    number  = "10",
    pages   = "103524",
    year    = "2019"
}

@article{Weinberg:1988cp,
    author  = "Weinberg, Steven",
    title   = "{The Cosmological Constant Problem}",
    journal = "Rev. Mod. Phys.",
    volume  = "61",
    pages   = "1--23",
    year    = "1989",
    doi     = "10.1103/RevModPhys.61.1"
}

@article{Peebles:2002gy,
    author  = "Peebles, P. J. E. and Ratra, Bharat",
    title   = "{The Cosmological Constant and Dark Energy}",
    eprint  = "astro-ph/0207347",
    archivePrefix = "arXiv",
    doi     = "10.1103/RevModPhys.75.559",
    journal = "Rev. Mod. Phys.",
    volume  = "75",
    pages   = "559--606",
    year    = "2003"
}

@article{Bamba:2012cp,
    author  = "Bamba, Kazuharu and Capozziello, Salvatore and Nojiri, Shin'ichi and Odintsov, Sergei D.",
    title   = "{Dark energy cosmology: the equivalent description via different theoretical models and cosmography tests}",
    eprint  = "1205.3421",
    archivePrefix = "arXiv",
    primaryClass  = "gr-qc",
    doi     = "10.1007/s10509-012-1181-8",
    journal = "Astrophys. Space Sci.",
    volume  = "342",
    pages   = "155--228",
    year    = "2012"
}

@article{Nojiri:2022nof,
    author  = "Nojiri, Shin'ichi and Odintsov, Sergei D. and Faraoni, Valerio",
    title   = "{From nonextensive statistics and black hole entropy to the holographic dark universe}",
    eprint  = "2201.02424",
    archivePrefix = "arXiv",
    primaryClass  = "gr-qc",
    doi     = "10.1103/PhysRevD.105.044042",
    journal = "Phys. Rev. D",
    volume  = "105",
    number  = "4",
    pages   = "044042",
    year    = "2022"
}

@article{Nojiri:2021jxf,
    author  = "Nojiri, Shin'ichi and Odintsov, Sergei D. and Faraoni, Valerio",
    title   = "{Area-law versus R\'enyi and Tsallis black hole entropies}",
    eprint  = "2109.05315",
    archivePrefix = "arXiv",
    primaryClass  = "gr-qc",
    doi     = "10.1103/PhysRevD.104.084030",
    journal = "Phys. Rev. D",
    volume  = "104",
    number  = "8",
    pages   = "084030",
    year    = "2021"
}

@article{Nojiri:2022sfd,
    author  = "Nojiri, Shin'ichi and Odintsov, Sergei D. and Faraoni, Valerio",
    title   = "{Alternative entropies and consistent black hole thermodynamics}",
    eprint  = "2207.07905",
    archivePrefix = "arXiv",
    primaryClass  = "gr-qc",
    doi     = "10.1142/S021988782250210X",
    journal = "Int. J. Geom. Meth. Mod. Phys.",
    volume  = "19",
    number  = "14",
    pages   = "2250210",
    year    = "2022"
}

@article{Nojiri:2019skr,
    author  = "Nojiri, Shin'ichi and Odintsov, Sergei D. and Saridakis, Emmanuel N.",
    title   = "{Modified cosmology from extended entropy with varying exponent}",
    eprint  = "1903.03098",
    archivePrefix = "arXiv",
    primaryClass  = "gr-qc",
    doi     = "10.1140/epjc/s10052-019-6740-5",
    journal = "Eur. Phys. J. C",
    volume  = "79",
    number  = "3",
    pages   = "242",
    year    = "2019"
}

@article{Nojiri:2020wmh,
    author  = "Nojiri, Shin'ichi and Odintsov, Sergei D. and Oikonomou, V. K. and Paul, Tanmoy",
    title   = "{Unifying holographic inflation with holographic dark energy: a covariant approach}",
    eprint  = "2007.06829",
    archivePrefix = "arXiv",
    primaryClass  = "gr-qc",
    doi     = "10.1103/PhysRevD.102.023540",
    journal = "Phys. Rev. D",
    volume  = "102",
    number  = "2",
    pages   = "023540",
    year    = "2020"
}

@article{Nojiri:2006ri,
    author  = "Nojiri, Shin'ichi and Odintsov, Sergei D.",
    title   = "{Introduction to modified gravity and gravitational alternative for dark energy}",
    eprint  = "hep-th/0601213",
    archivePrefix = "arXiv",
    doi     = "10.1142/S0219887807001928",
    journal = "Int. J. Geom. Meth. Mod. Phys.",
    volume  = "4",
    pages   = "115--146",
    year    = "2007"
}

@article{Nojiri:2017ncd,
    author  = "Nojiri, Shin'ichi and Odintsov, Sergei D. and Oikonomou, V. K.",
    title   = "{Modified Gravity Theories on a Nutshell: Inflation, Bounce and Late-time Evolution}",
    eprint  = "1705.11098",
    archivePrefix = "arXiv",
    primaryClass  = "gr-qc",
    doi     = "10.1016/j.physrep.2017.06.001",
    journal = "Phys. Rept.",
    volume  = "692",
    pages   = "1--104",
    year    = "2017"
}

@article{Capozziello:2011et,
    author  = "Capozziello, Salvatore and De Laurentis, Mariafelicia",
    title   = "{Extended Theories of Gravity}",
    eprint  = "1108.6266",
    archivePrefix = "arXiv",
    primaryClass  = "gr-qc",
    doi     = "10.1016/j.physrep.2011.09.003",
    journal = "Phys. Rept.",
    volume  = "509",
    pages   = "167--321",
    year    = "2011"
}

@article{Riess:2021jrx,
    author  = "Riess, Adam G. and others",
    title   = "{A Comprehensive Measurement of the Local Value of the Hubble Constant with 1 km s$^{-1}$ Mpc$^{-1}$ Uncertainty from the Hubble Space Telescope and the SH0ES Team}",
    eprint  = "2112.04510",
    archivePrefix = "arXiv",
    primaryClass  = "astro-ph.CO",
    doi     = "10.3847/2041-8213/ac5c5b",
    journal = "Astrophys. J. Lett.",
    volume  = "934",
    number  = "1",
    pages   = "L7",
    year    = "2022"
}

@article{DiValentino:2021izs,
    author  = "Di Valentino, Eleonora and Mena, Olga and Pan, Supriya and Visinelli, Luca and Yang, Weiqiang and Melchiorri, Alessandro and Mota, David F. and Riess, Adam G. and Silk, Joseph",
    title   = "{In the realm of the Hubble tension -- a review of solutions}",
    eprint  = "2103.01183",
    archivePrefix = "arXiv",
    primaryClass  = "astro-ph.CO",
    doi     = "10.1088/1361-6382/ac086d",
    journal = "Class. Quant. Grav.",
    volume  = "38",
    number  = "15",
    pages   = "153001",
    year    = "2021"
}

@article{DESI:2024mwx,
    author  = "Adame, A. G. and others",
    collaboration = "DESI",
    title   = "{DESI 2024 VI: cosmological constraints from the measurements of baryon acoustic oscillations}",
    eprint  = "2404.03002",
    archivePrefix = "arXiv",
    primaryClass  = "astro-ph.CO",
    doi     = "10.1088/1475-7516/2025/02/021",
    journal = "JCAP",
    volume  = "02",
    pages   = "021",
    year    = "2025"
}

@article{Huang:2004ai,
    author  = "Huang, Qing-Guo and Li, Miao",
    title   = "{The holographic dark energy in a non-flat universe}",
    eprint  = "astro-ph/0404229",
    archivePrefix = "arXiv",
    doi     = "10.1088/1475-7516/2004/08/013",
    journal = "JCAP",
    volume  = "08",
    pages   = "013",
    year    = "2004"
}

@article{Pavon:2005yx,
    author  = "Pav\'on, Diego and Zimdahl, Winfried",
    title   = "{Holographic dark energy and cosmic coincidence}",
    eprint  = "gr-qc/0505020",
    archivePrefix = "arXiv",
    doi     = "10.1016/j.physletb.2005.08.134",
    journal = "Phys. Lett. B",
    volume  = "628",
    pages   = "206--210",
    year    = "2005"
}

@article{Elizalde:2005ju,
    author  = "Elizalde, Emilio and Nojiri, Shin'ichi and Odintsov, Sergei D. and Wang, Peng",
    title   = "{Dark energy: Vacuum fluctuations, the effective phantom phase, and holography}",
    eprint  = "hep-th/0502082",
    archivePrefix = "arXiv",
    doi     = "10.1103/PhysRevD.71.103504",
    journal = "Phys. Rev. D",
    volume  = "71",
    pages   = "103504",
    year    = "2005"
}

@article{Bekenstein:1973ur,
    author  = "Bekenstein, Jacob D.",
    title   = "{Black holes and entropy}",
    doi     = "10.1103/PhysRevD.7.2333",
    journal = "Phys. Rev. D",
    volume  = "7",
    pages   = "2333--2346",
    year    = "1973"
}

@article{Hawking:1975vcx,
    author  = "Hawking, S. W.",
    title   = "{Particle Creation by Black Holes}",
    doi     = "10.1007/BF02345020",
    journal = "Commun. Math. Phys.",
    volume  = "43",
    pages   = "199--220",
    year    = "1975",
    note    = "[Erratum: Commun. Math. Phys. 46, 206 (1976)]"
}

@article{Tsallis:1987eu,
    author  = "Tsallis, Constantino",
    title   = "{Possible Generalization of Boltzmann-Gibbs Statistics}",
    doi     = "10.1007/BF01016429",
    journal = "J. Statist. Phys.",
    volume  = "52",
    pages   = "479--487",
    year    = "1988"
}

@article{Tavayef:2018xwx,
    author  = "Tavayef, M. and Sheykhi, A. and Bamba, Kazuharu and Moradpour, H.",
    title   = "{Tsallis Holographic Dark Energy}",
    eprint  = "1804.02983",
    archivePrefix = "arXiv",
    primaryClass  = "gr-qc",
    doi     = "10.1016/j.physletb.2018.04.001",
    journal = "Phys. Lett. B",
    volume  = "781",
    pages   = "195--200",
    year    = "2018"
}

@article{Moradpour:2018ivi,
    author  = "Moradpour, H. and Moosavi, S. A. and Lobo, I. P. and Morais Gra\c{c}a, J. P. and Jawad, A. and Salako, I. G.",
    title   = "{Thermodynamic approach to holographic dark energy and the R\'enyi entropy}",
    eprint  = "1803.02195",
    archivePrefix = "arXiv",
    primaryClass  = "physics.gen-ph",
    doi     = "10.1140/epjc/s10052-018-6309-8",
    journal = "Eur. Phys. J. C",
    volume  = "78",
    number  = "10",
    pages   = "829",
    year    = "2018"
}

@article{Drepanou:2021jiv,
    author  = "Drepanou, Niki and Lymperis, Andreas and Saridakis, Emmanuel N. and Yesmakhanova, Kuralay",
    title   = "{Kaniadakis holographic dark energy and cosmology}",
    eprint  = "2109.09181",
    archivePrefix = "arXiv",
    primaryClass  = "gr-qc",
    doi     = "10.1140/epjc/s10052-022-10415-9",
    journal = "Eur. Phys. J. C",
    volume  = "82",
    number  = "5",
    pages   = "449",
    year    = "2022"
}

@article{Nojiri:2024eac,
    author  = "Nojiri, Shin'ichi and Odintsov, Sergei D. and Paul, Tanmoy",
    title   = "{Different Aspects of Entropic Cosmology}",
    eprint  = "2409.01090",
    archivePrefix = "arXiv",
    primaryClass  = "gr-qc",
    doi     = "10.3390/universe10090352",
    journal = "Universe",
    volume  = "10",
    number  = "9",
    pages   = "352",
    year    = "2024"
}

@misc{Saridakis:2026rev,
    author  = "Saridakis, Emmanuel N. and Luciano, Giuseppe Gaetano and Lymperis, Andreas and Telali, Eleni C.",
    title   = "{Entropy applications in cosmology: spacetime thermodynamics, holographic dark energy, entropic gravity and beyond - a review}",
    eprint  = "2609.28331",
    archivePrefix = "arXiv",
    primaryClass  = "astro-ph.CO",
    year    = "2026"
}

@article{Sharma:1975,
    author  = "Sharma, B. D. and Mittal, D. P.",
    title   = "{New non-additive measures of entropy for discrete probability distributions}",
    journal = "J. Math. Sci.",
    volume  = "10",
    pages   = "28--40",
    year    = "1975"
}

@article{Chevallier:2000qy,
    author = "Chevallier, Michel and Polarski, David",
    title = "{Accelerating universes with scaling dark matter}",
    eprint = "gr-qc/0009008",
    archivePrefix = "arXiv",
    doi = "10.1142/S0218271801000822",
    journal = "Int. J. Mod. Phys. D",
    volume = "10",
    pages = "213--224",
    year = "2001"
}

@article{Linder:2002et,
    author = "Linder, Eric V.",
    title = "{Exploring the expansion history of the universe}",
    eprint = "astro-ph/0208512",
    archivePrefix = "arXiv",
    doi = "10.1103/PhysRevLett.90.091301",
    journal = "Phys. Rev. Lett.",
    volume = "90",
    pages = "091301",
    year = "2003"
}

\end{document}